\documentstyle[11pt,amsmath,amssymb]{article}

\renewcommand{\thesection}{\arabic{section}}

\newtheorem{theorem}{Theorem}[section]

\newtheorem{proposition}[theorem]{Proposition}
\newtheorem{lemma}[theorem]{Lemma}
\newtheorem{corollary}[theorem]{Corollary}

\renewcommand{\d}{{\rm d}}
\newcommand{\bbbone}{\mathchoice {\rm 1\mskip-4mu l} {\rm 1\mskip-4mu l}
{\rm 1\mskip-4.5mu l} {\rm 1\mskip-5mu l}}

\newcommand{\scalprod}[2]{\left\langle {#1}, {#2}\right\rangle}
\newcommand{\fer}[1]{(\ref{#1})}
\newcommand{\av}[1]{\left\langle{#1}\right\rangle}

\newcommand{\s}{{\rm S}}
\newcommand{\h}{{\cal H}}
\newcommand{\hh}{{\frak h}}
\newcommand{\cx}{{\mathbb C}}
\newcommand{\rhobar}{{\overline \rho}}
\newcommand{\e}{{\,\rm e}}
\renewcommand{\i}{{\rm i}}
\newcommand{\rx}{{\mathbb R}}
\newcommand{\tr}{{\rm Tr}}
\newcommand{\r}{{\rm R}}
\newcommand{\mm}{{\frak M}}
\newcommand{\cc}{{\cal C}}

\renewcommand{\theequation}{\arabic{equation}}

\let\sect\section
\renewcommand\section{\setcounter{equation}{0}
\gdef\theequation{\thesection.\arabic{equation}}\sect}

\begin{document}
\title{Resonance Theory of Decoherence and Thermalization}

\author{
M. Merkli$^{\rm 1}$\footnote{Email: merkli@math.mun.ca; Partly supported by NSERC under grant NA 7901.}\ , \ \ \ I.M. Sigal$^{\rm 2}$\footnote{Email: im.sigal@utoronto.ca; Supported by NSERC under grant NA 7901.}\ ,\ \ \ G.P. Berman$^{\rm 3}$\footnote{Email: gpb@lanl.gov; Supported by the NNSA of the U.S. DOE at LANL under Contract No. DE-AC52-06NA25396.}
\vspace*{.3cm}
\\
$^{\rm 1}${\it Department of Mathematics and Statistics,}\\
{\it  Memorial University of Newfoundland}\\
\smallskip {\it  St. John's, Newfoundland, Canada, A1C 5S7}\\
$^{\rm 2}${\it Department of Mathematics, University of Toronto}\\
\smallskip {\it Toronto, Ontario, Canada, M5S 2E4}\\
$^{\rm 3}${\it Theoretical Division, MS B213, Los Alamos National Laboratory} \\
{\it Los Alamos, NM 87545, USA}\\
}
\date{\today}
\maketitle

\begin{abstract}
We present a rigorous analysis of the phenomenon of decoherence
for general $N-$level systems coupled to reservoirs. The
latter are described by free massless bosonic fields. We apply our
general results to the specific cases of the qubit and the quantum
register. We compare our results with the explicitly solvable case
of systems whose interaction with the environment does not allow for
energy exchange (non-demolition, or energy conserving interactions). We suggest a new approach which applies to a wide variety of systems which are not explicitly solvable. 
\end{abstract}

\setlength{\baselineskip}{15pt}

\section{Introduction}
\label{sectintrodd}

In this paper we examine rigorously the phenomenon of quantum decoherence. This phenomenon is brought about
by the interaction of a quantum system, called  here ``the system $\s$'' for short, with an environment, see e.g. \cite{JZKGKS,SH,Z} and the many references therein. 
Decoherence is reflected in the temporal decay of off-diagonal elements of the reduced density matrix of the
system in a given basis. So far, this phenomenon has been analyzed rigorously only for explicitly solvable
models, \cite{DBV,DG,JZKGKS,MP,SH,SGC,PSE,VK,Z}. In this paper we consider the decoherence phenomenon for quite
general non-solvable models. Our analysis is based on the modern theory of resonances for quantum statistical
systems as developed in \cite{ASF,BFS0,JP1,BFS,JP2,HuS,SV,MMS1,MMS2} (see also the book \cite{HiS}), which is
related to resonance theory in non-relativistic quantum electrodynamics (\cite{BFS,BCFS}).

Let $\hh=\hh_\s\otimes\hh_\r$ be the Hilbert space of the system
interacting with the environment, also called the ``reservoir'', and let
\begin{equation}
H= H_\s\otimes\bbbone_\r + \bbbone_\s\otimes H_\r +\lambda v
\label{ii1}
\end{equation}
be its Hamiltonian. Here, $H_\s$ and $H_\r$ are the Hamiltonians of the system and the reservoir, respectively, and $\lambda v$ is an interaction with a coupling constant $\lambda\in\rx$. We will omit trivial factors $\bbbone_\s\otimes$ and  $\otimes\bbbone_\r$ when there is no danger of confusion. In this paper we consider finite dimensional systems, the reservoirs are described by free massless quantum fields, and we take interactions of the form $v=G\otimes\varphi(g)$, 
where $G$ is a self-adjoint matrix on $\hh_\s$ and $\varphi(g)$ is the field operator, smoothed out with a coupling function (form factor).

Consider possibly entangled initial states of the system and reservoir, where the reservoir is close to (a local perturbation of) an equilibrium state at some temperature $T=1/\beta>0$. (In the literature on decoherence, most often it is assumed that the initial states are product states, where the reservoir is in equilibrium, but our method works in the general case.)  Let $\rho_t$ be the density matrix of the total system at time $t$. The reduced density matrix (of the system $\s$) at time $t$ is then formally given by
$$
\rhobar_t = \tr_\r\,\rho_t,
$$
where $\tr_\r$ is the partial trace with respect to the reservoir degrees of freedom. For the sake of explicitness we describe here the case where the state of the reservoir is given by a well-defined density matrix on the Hilbert space $\hh_\r$. In the next section we define the relevant notions for a more realistic reservoir, obtained for instance by taking a thermodynamic limit, or a continuous-mode limit.

Let $\rho(\beta,\lambda)$ be the equilibrium state of the interacting system at temperature $T=1/\beta$ and set $\rhobar(\beta,\lambda):=\tr_\r\rho(\beta,\lambda)$. There are three possible scenarios for the asymptotic behaviour of the reduced density matrix, as $t\rightarrow\infty$:
\begin{itemize}
\item[(i)]  $\rhobar_t\longrightarrow \rhobar_\infty = \rhobar(\beta,\lambda)$,
\item[(ii)] $\rhobar_t\longrightarrow \rhobar_\infty \neq  \rhobar(\beta,\lambda)$,
\item[(iii)] $\rhobar_t$ does not converge.
\end{itemize}
The first situation is generic while the last two are not, although they are of interest, e.g. for energy conserving, or quantum non-demolition interactions, characterized by $[H_\s,v]=0$, see \cite{JZKGKS} and Section \ref{sectapp2}.

Decoherence is a basis-dependent notion. It is usually defined as the vanishing of the off-diagonal elements $[\rhobar_t]_{m,n}$, $m\neq n$ in the limit $t\rightarrow\infty$, in a chosen basis. Most often decoherence is defined w.r.t. the basis of eigenvectors of the system Hamiltonian $H_\s$ (the energy basis, also called the computational basis for a quantum register), though other bases, such as the position basis for a particle in a scattering medium \cite{JZKGKS}, are also used.

Since $\rhobar(\beta,\lambda)$ is generically non-diagonal in the energy basis, the off-diagonal elements of $\rhobar_t$ will not vanish in the generic case, as $t\rightarrow\infty$. Thus, strictly speaking, decoherence in this case should be defined as the decay (convergence) of the off-diagonals of $\rhobar_t$ to the corresponding off-diagonals of $\rhobar(\beta,\lambda)$. The latter are of the order $O(\lambda)$ and in concrete applications often of the order $O(\lambda^2)$.
If these terms are neglected, then decoherence manifests itself as a process in which initially coherent superpositions of basis elements $\psi_j$ become incoherent statistical mixtures,
$$
\sum_{j,k}c_{j,k}|\psi_j\rangle\langle\psi_k| \longrightarrow \sum_j p_j |\psi_j\rangle\langle\psi_j|,\ \ \ \mbox{ as $t\rightarrow\infty$}.
$$
In particular, phase relations encoded in the $c_{j,k}$, $j\neq k$, disappear
for large times. Of course, as $t\rightarrow\infty$, the
off-diagonal elements of $\rhobar_t$ vanish in a basis of
eigenvectors of the asymptotic Hamiltonian
$H_{\s,\lambda,\beta}:=-\frac 1\beta \ln
\rhobar(\beta,\lambda)=H_\s+O(\lambda)$. We conjecture that this
Hamiltonian absorbs the leading order correction to the
non-interacting dynamics $\e^{-\i t H_\s}$ due to the interaction with
the reservoir. 
We discuss the role of $H_{\s,\lambda,\beta}$ in more
detail elsewhere. We set $\hbar$ equal to one in what follows.

\bigskip
In this paper we consider $N$-dimensional quantum systems interacting in a quite general way with reservoirs of massless free quantum fields (photons, phonons or other massless excitations). Let $A$ be an arbitrary observable of the system (an operator on the system Hilbert space $\hh_\s$) and set
\begin{equation}
\av{A}_t := \tr_\s(\rhobar_t A) =\tr_{\s +\r}(\rho_t (A\otimes\bbbone_\r) ).
\label{avA}
\end{equation}
We show, under certain conditions on the interaction, that the ergodic averages
\begin{equation}
\av{\av{A}}_\infty := \lim_{T\rightarrow\infty}\frac 1T\int_0^T\av{A}_t \d t
\label{ergav}
\end{equation}
exist, i.e., that $\av{A}_t$ converges in the ergodic sense as $t\rightarrow\infty$. Furthermore, we show that for any $t\geq 0$ and for any $0<\omega'<\frac{2\pi}{\beta}$,
\begin{equation}
\av{A}_t-\av{\av{A}}_\infty =\sum_{\varepsilon \neq 0}\e^{\i t\varepsilon} R_\varepsilon(A) +O\left(\lambda^2\e^{-\frac t2 [\max_\varepsilon\{{\rm Im}\, \varepsilon\} + \omega'/2]}\right),
\label{introdd1}
\end{equation}
where the complex numbers $\varepsilon$ are the  eigenvalues of a certain explicitly given operator $K(\omega')$, lying in the strip $\{z\in\cx\ |\ 0\leq {\mathrm Im} z<\omega'/2\}$.  They have the expansions
\begin{equation}
\varepsilon\equiv \varepsilon_e^{(s)} = e -\lambda^2\delta_e^{(s)} + O(\lambda^4),
\label{int1}
\end{equation}
where $e\in{\mathrm spec}(H_\s\otimes\bbbone_\s-\bbbone_\s\otimes
H_\s) = {\mathrm spec}(H_\s)- {\mathrm spec}(H_\s)$ and the 
$\delta_e^{(s)}$ are the eigenvalues of a matrix $\Lambda_e$, called
a {\it level-shift operator}, acting on the eigenspace of
$H_\s\otimes\bbbone_\s-\bbbone_\s\otimes H_\s$ corresponding to the
eigenvalue $e$ (which is a subspace of $\hh_\s\otimes\hh_\s$). The
level shift operators play a central role in the ergodic theory of
open quantum systems, see e.g. \cite{Mlso}. We describe them in
Section \ref{dynrestheory}. The terms $R_\varepsilon(A)$ in
\fer{introdd1} are linear functionals of $A$ and are given in terms
of the initial state, $\rho_0$, and certain operators depending on
the Hamiltonian $H$. They have the expansion
$$
R_\varepsilon(A)=\sum_{(m,n)\in I_e}\varkappa_{m,n} A_{m,n} +O(\lambda^2),
$$
where $I_e$ is the collection of all pairs of indices such that $e=E_m-E_n$, the $E_k$ being the eigenvalues of $H_\s$. Here, $A_{m,n}$ is the $(m,n)$-matrix element of the observable $A$ in the energy basis of $H_\s$, and the $\varkappa_{m,n}$ are coefficients depending on the initial state of the system (and on $e$, but not on $A$ nor on $\lambda$).

\smallskip
{\it Discussion of \fer{introdd1}.\ } In the absence of interaction
($\lambda=0$) we have $\varepsilon=e\in{\mathbb R}$, see \fer{int1}.
Depending on the interaction each resonance energy $\varepsilon$ may
migrate into the upper complex plane, or it may stay on the real
axis, as $\lambda\neq 0$. The averages $\av{A}_t$ approach their
ergodic means $\av{\av{A}}_\infty$ if and only if $\mathrm Im
\varepsilon
>0$ for all $\varepsilon\neq 0$. In this case the convergence takes place
on the time scale $[{\mathrm Im}\varepsilon]^{-1}$. Otherwise
$\av{A}_t$ oscillates.
A sufficient condition for decay is that ${\rm Im}\delta_e^{(s)} <0$
(and $\lambda$ small, see \fer{int1}).

{\it Remark about the error term in \fer{introdd1}.\ } The restrictive condition ${\rm Im}\,\varepsilon<\omega'/2<\pi/\beta$ in \fer{introdd1} which implies $\beta\leq c\lambda^2$, for some constant $c$, can be eliminated by using renormalization group methods as in \cite{BFS,Mthesis}, and our results can be upgraded to hold uniformly in $T=1/\beta\rightarrow 0$. This point will be addressed elsewhere. 

There are two kinds of processes which drive the decay:
energy-exchange processes and energy preserving ones. The former are
induced by interactions having nonvanishing probabilities for
processes of absorption and emission of field quanta with energies
corresponding to the Bohr frequencies of $\s$ (this is the ``Fermi
Golden Rule Condition'', \cite{BFS,FM1,Mlso,MMS1,MMS2}). Energy
preserving interactions suppress such processes, allowing only for a
phase change of the system during the evolution (``phase damping'',
\cite{PSE,BBD,DBV,DG,JZKGKS,MP,SGC}).

Relation \fer{introdd1} gives a detailed picture of the dynamics of
averages of observables. The resonance energies $\varepsilon$ and
the functionals $R_\varepsilon$ can be calculated for concrete
models, to arbitrary precision (in the sense of rigorous perturbation theory
in $\lambda$). See \fer{i1}-\fer{i4} for explicit expressions for
the qubit, and the illustration below for an initially coherent
superposition given by \fer{illust}. In this paper we use relation
\fer{introdd1} to discuss the processes of thermalization and
decoherence. It would be interesting to apply the techniques
developed here to the analysis of the transition from quantum
behaviour to classical behaviour (see \cite{BBBD,DBV}).

\medskip

We apply our results to a qubit, as well as to energy-preserving, or non-demolition interactions. They apply
equally well to a register of arbitrarily many qubits. The case of energy-preserving interactions can be
solved explicitly and serves as an illustrative example as well as a starting point for a perturbation theory
for interactions which are not energy-preserving, but for which the commutator $[H_\s,v]$ is small.

\medskip
Our results for the qubit can be summarized as follows. Consider a
qubit coupled linearly to the field by the interaction
\begin{equation}
v = \left[
\begin{array}{cc}
 a & c\\
\overline c & b
\end{array}
\right]
\otimes\varphi(g),
\label{michael*}
\end{equation}
where $\varphi(g)$ is the Bose field operator, smeared out with a
coupling function (form factor) $g(k)$, $k\in\rx^3$, and the
$2\times 2$ coupling matrix (representing the coupling operator in
the energy eigenbasis) is hermitian. The operator \fer{michael*} -
or a sum of such terms, for which our technique works equally well -
is the most general coupling which is linear in field operators. We refer to Remark 14 below for the link between \fer{michael*} and the spin-boson model.

Note that the form-factor $g$ contains an ultra-violet cut-off which
introduces a time-scale $\tau_{UV}$. This time scale depends on the
physical system in question. We can think of it as coming from some frequency-cutoff determined by a characteristic length scale beyond which the interaction decreases rapidly. For instance, for a phonon field $\tau_{UV}$ is naturally identified with the inverse of the Debye frequency.
We assume $\tau_{UV}$ to be much smaller than the time scales considered here.

A key role in the decoherence analysis is played by the infrared
behaviour of form factors $g\in L^2(\rx^3,\d^3k)$. We characterize
this behavior by the unique $p\geq -1/2$ satisfying
\begin{equation}
0< \lim_{|k|\rightarrow 0} \frac{|g(k)|}{|k|^p}=C<\infty. \label{38}
\end{equation}
The power $p$ depends on the physical model considered, e.g.
for quantum-optical systems, $p=1/2$, and for the quantized electromagnetic field, $p=-1/2$.

Decoherence of models with interaction \fer{michael*} with $c=0$ is
considered in \cite{BBD,DBV,DG,JZKGKS,MP,PSE,SGC,U} (see also
Section \ref{sectapp2}). This is the  situation of a non-demolition
(energy conserving) interaction, where $v$ commutes with the
Hamiltonian $H_\s$ and consequently energy-exchange processes are
suppressed. The resulting decoherence is called phase-decoherence. A
particular model of phase-decoherence is obtained by the so-called
position-position coupling, where the matrix in the interaction
\fer{michael*} is the Pauli matrix $\sigma_z$ \cite{BBD,DG,PSE,U}.
On the other hand,
energy-exchange processes, responsible for driving the system to
equilibrium, have a probability proportional to $|c|^{2n}$, for some
$n\geq 1$ (and $a$, $b$ do not enter)
\cite{ASF2,BFS,FM1,JP1,Mlso,MMS1,MMS2}. Thus the property $c\neq 0$ is
important for thermalization (return to equilibrium).

We express the energy-exchange effectiveness in terms of the function
\begin{equation}
\xi(\eta) = \lim_{\epsilon\downarrow 0} \frac 1\pi\int_{\rx^3}\d^3k\coth\!\left(\frac{\beta |k|}{2}\right)
 |g(k)|^2 \frac{\epsilon}{(|k|-\eta)^2+\epsilon^2},
\label{35}
\end{equation}
where $\eta\geq 0$ represents the energy at which processes between
the qubit and the reservoir take place. Let $\Delta=E_2-E_1>0$ be
the energy gap of the qubit. In works on convergence to equilibrium
it is usually assumed that $|c|^2\xi(\Delta)>0$. This condition is
called the ``Fermi Golden Rule Condition''. It means that the
interaction induces second-order ($\lambda^2$) energy exchanging
processes at the Bohr frequency of the qubit (emission and
absorption of reservoir quanta). The condition $c\neq 0$ is actually
{\it necessary} for thermalization while $\xi(\Delta)>0$ is not
(higher order processes can drive the system to equilibrium).
Observe that $\xi(\Delta)$ converges to a fixed function, $\xi_0(\Delta)$, as $T\rightarrow 0$, and $\xi(\Delta)$ increases exponentially as
$T\rightarrow \infty$. The expression for the decoherence involves
also $\xi(0)$ (see below).

In this paper we describe the dynamics, and in particular the
decoherence properties, of systems which exhibit {\it both}
thermalization {\it and} (phase) decoherence. See the the discussion
after \fer{38} for a comparison of the two effects, and how they
relate to the coefficients $a,b,c$ and the coupling function $g$ in
\fer{michael*}.

Let the initial density matrix, $\rho_{t=0}$, be of the form
$\rhobar_0\otimes\rho_{\r,\beta}$. (Our method does not require the
initial state to be a product, see Remark 5 below.) Denote by
$p_{m,n}$ the rank-one operator represented in the energy
basis by the $2\times 2$ matrix whose entries are zero, except the
$(n,m)$ entry which is one. We show that for $t\geq 0$
\begin{eqnarray}
[\rhobar_t]_{1,1} -\av{\av{p_{1,1}}}_\infty &=& \e^{\i
t\varepsilon_0(\lambda)} \left[ C_0
+O(\lambda^2)\right]\label{i1}\\
&&+ \e^{\i t\varepsilon_\Delta(\lambda)} O(\lambda^2) + \e^{\i t\varepsilon_{-\Delta}(\lambda)} O(\lambda^2)
+ O(\lambda^2\e^{-t\omega'/2})
\nonumber
\end{eqnarray}
and
\begin{eqnarray}
[\rhobar_t]_{1,2} -\av{\av{p_{1,2}}}_\infty &=& \e^{\i
t\varepsilon_\Delta(\lambda)} \left[C_\Delta
+O(\lambda^2)\right]\label{i2}\\
&&+ \e^{\i t\varepsilon_0(\lambda)} O(\lambda^2) + \e^{\i t\varepsilon_{-\Delta}(\lambda)} O(\lambda^2)
+ O(\lambda^2\e^{-t\omega'/2}).
\nonumber
\end{eqnarray}
Here, $C_0, C_\Delta$ are explicit constants depending on the initial condition $\rhobar_0$, but not on $\lambda$, and the resonance energies $\varepsilon$ have the expansions
\begin{eqnarray}
\varepsilon_0(\lambda) &=& \i\lambda^2 \pi^2|c|^2\xi(\Delta) +O(\lambda^4)\nonumber\\
\varepsilon_\Delta(\lambda) &=& \Delta +\lambda^2 R +{\textstyle \frac \i2}\lambda^2\pi^2\left[ |c|^2 \xi
(\Delta)+(b-a)^2\xi(0)\right]  +O(\lambda^4)\label{i4}\\
\varepsilon_{-\Delta}(\lambda) &=& -\overline{\varepsilon_\Delta(\lambda)}\nonumber
\end{eqnarray}
with the real number
$$
R = {\textstyle \frac 12}(b^2-a^2) \scalprod{g}{\omega^{-1}g}
$$
$$+{\textstyle \frac 12}|c|^2 {\rm P.V.}\int_{\rx\times S^2} u^2|g(|u|,\sigma)|^2 \coth\!\left(\frac{\beta |u|}{2}\right)\frac{1}{u-\Delta}.$$
The error terms in \fer{i1}, \fer{i2} and \fer{i4} satisfy, for
small $\lambda$,
$$
\left|\frac{O(\lambda^2)}{\lambda^{2}}\right|<C \mbox{\ \ and\ \ }
\sup_{t\geq
0}\left|\frac{O(\lambda^2\e^{-t\omega'/2})}{\lambda^{2}\e^{-t\omega'/2}}\right|<C.
$$

{\it Remarks.\ }
1)\ To our knowledge this is the first time that formulas \fer{i1}-\fer{i4} are presented for models which are not explicitly solvable. Results for exactly solvable models (non-demolition interactions) are given, among others, in \cite{PSE,SGC,MP}. See \cite{JZKGKS} for an overview of the subject.

2)\ Relations \fer{i1}-\fer{i4} are valid {\it for all values of $t\geq 0$}, and the remainder terms $O(\lambda^2)$ are uniform in $t\geq 0$. In particular, we do {\it not} require that $\lambda\rightarrow 0$ as $t\rightarrow\infty$ (van Hove limit).

3)\ Even if the initial density matrix,
$\rho_{t=0}$, is a product of the system and reservoir density
matrices, the density matrix, $\rho_t$, at any subsequent moment of
time $t>0$ is not of the product form. In other words, the evolution
creates the system-reservoir entanglement.

4)\ The corresponding expressions for the matrix elements
$[\rhobar_t]_{2,2}$ and $[\rhobar_t]_{2,1}$ are obtained from the
relations $[\rhobar_t]_{2,2}=1-[\rhobar_t]_{1,1}$ (conservation of
unit trace) and $[\rhobar_t]_{2,1} = [\rhobar_t]_{1,2}^*$
(hermiticity of $\rhobar_t$).

5)\ If the qubit is initially in one of the logic pure states
$\rhobar_0=|\varphi_j\rangle\langle\varphi_j|$, where
$H_\s\varphi_j=E_j\varphi_j$, $j=1,2$, then  we have $C_\Delta=0$,
and $C_0=\e^{\beta\Delta/2}(\e^{\beta\Delta}+1)^{-3/2}$ for $j=1$
and $C_0=\e^{\beta\Delta}(\e^{\beta\Delta}+1)^{-3/2}$ for $j=2$, see
at the end of Section \ref{subsqubit}.

6)\ We develop a formula for $\av{A}_t-\av{\av{A}}_\infty$ for all
observables $A$ of any $N$-level system $\s$ in Section
\ref{dynrestheory}.

7)\ If the system has the property of return to equilibrium, i.e.,
if $\xi(\Delta)>0$, then
$$
\av{\av{p_{n,m}}}= [\rhobar_\infty]_{m,n} = \delta_{m,n}\frac{e^{-\beta E_m}}{\tr_\s(\e^{-\beta H_\s})} + O(\lambda^2).
$$
We thus recover the Gibbs law in the long time limit followed by the weak coupling limit. A similar observation is found in the context of the quantum Langevin equation in \cite{BK}. 

8)\ If $\rho_0$ is an arbitrary initial density matrix on
$\h_\s\otimes\h_\r$ (i.e., not necessarily of product form), then the method developed in Section
\ref{dynrestheory} yields the following result: For any $\eta>0$
there are constants $C_0$, $C_\Delta$, depending on $\eta$ and
$\rho_0$ but not on $\lambda$, s.t.
\begin{eqnarray}
[\rhobar_t]_{1,1} -\av{\av{p_{1,1}}}_\infty &=& \e^{\i
t\varepsilon_0(\lambda)} \left[ C_0
+O(\lambda)\right]\label{i11}\\
&&+ \e^{i t\varepsilon_\Delta(\lambda)} O(\lambda) + \e^{i t\varepsilon_{-\Delta}(\lambda)} O(\lambda) + O(\lambda \e^{-t\omega'/2}) +O(\eta)
\nonumber
\end{eqnarray}
and
\begin{eqnarray}
[\rhobar_t]_{1,2} -\av{\av{p_{1,2}}}_\infty &=& \e^{\i
t\varepsilon_\Delta(\lambda)} \left[C_\Delta
+O(\lambda)\right]\label{i22}\\
&&+ \e^{i t\varepsilon_0(\lambda)} O(\lambda) + \e^{i t\varepsilon_{-\Delta}(\lambda)} O(\lambda) + O(\lambda \e^{-t\omega'/2})+O(\eta),\nonumber
\end{eqnarray}
where $O(\eta)$ is uniform in $t$, and where the resonance energies are given by \fer{i4}.
Furthermore, all remainder terms depend on $\eta$, in general.

9)\ Equations \fer{i1}, \fer{i2} and \fer{i4} define the decoherence time
scale,  $\tau_D=[{\mathrm Im}\varepsilon_\Delta(\lambda)]^{-1}$, and
the thermalization time scale,  $\tau_T=[{\mathrm
Im}\varepsilon_0(\lambda)]^{-1}$. We should compare
$\tau_D$ with the
decoherence time scales in real systems and with computational time
scales. The former depends on the physical realization of the qubit
and its environment. It can vary from $10^4$s for nuclear spins in
paramagnetic atoms to $10^{-12}$s for electron-hole excitations in
bulk semiconductors (see e.g. \cite{DiV}). ${\rm Re}\,\varepsilon_\Delta(\lambda)-\Delta=\lambda^2R +O(\lambda^4)$ gives the radiative energy shifts.

10)\ To second order in $\lambda$, the imaginary part of
$\varepsilon_\Delta$  is increased by a term $\propto (b-a)^2\xi(0)$
only if $p=-1/2$, where $p$ is defined in \fer{38}. For $p>-1/2$ we
have $\xi(0)=0$ and that contribution vanishes. For $p<-1/2$ we have
$\xi(0)=\infty$.

11)\ $\xi(\Delta)$ and $R$ contain purely quantum, vacuum fluctuation terms,  as well as thermal ones, while $\xi(0)$
is determined entirely by thermal fluctuations. $\xi(\Delta)$ and $\xi(0)$ are increasing in $T$, and, as $T\rightarrow 0$, $\xi(0)$ is linear in $T$ ($p=-1/2$) and $\xi(\Delta)$ converges to a fixed nonzero value. The decoherence rate thus increases for decreasing $T$, and it approaches a finite value as $T\rightarrow 0$, for $c\neq 0$. Our proofs work for arbitrarily small, fixed temperatures. There is strong evidence that the results above remain valid for $T\rightarrow 0$ (see the remark about the error term in \fer{introdd1} above, and also \cite{BFS,Mthesis}). A discussion of the decoherence function in terms of the temperature for the explicitly solvable case, $c=0$, is given in \cite{PSE}.

12)\  To second order in pertrubation, the ratio of the thermalization versus decoherence rate is $\tau_{\rm T}/\tau_{\rm D} = \frac{1}{2}[1+(\frac{b-a}{|c|})^2\frac{\xi(0)}{\xi(\Delta)}]$. For $\tau_{\rm T}/\tau_{\rm D}<1$, the populations converge to their limiting values faster than the off-digaonal matrix elements, as $t\rightarrow\infty$ (coherence persists beyond thermalization of the population). For $\tau_{\rm T}/\tau_{\rm D} >1$, the off-diagonal elements converge faster. If the interaction matrix is diagonal ($c=0$), then $\tau_{\rm T}/\tau_{\rm D} =\infty$, if it is off-diagonal (or if $a=b$), then $\tau_{\rm T}/\tau_{\rm D} =1/2$.

13)\ For energy-conserving interactions, $c=0$, it follows that full
decoherence occurs if and only if $b\neq a$ and $\xi(0)>0$. If
either of these conditions are not satisfied then the off-diagonal
matrix elements are purely oscillatory (while the populations are
constant). We analyze energy-conserving interactions in Section \ref{sectapp2}.

14)\ In the ubiquitous spin-boson model \cite{LCDFGZ}, obtained as a two-state truncation of a double-well system or an atom, interacting with a Bose field, the Hamiltonian is given by \fer{ii1} with $H_\s = -\frac 12\Delta_0\sigma_x +\frac 12\epsilon\sigma_z$ and $v=\sigma_z\otimes\varphi(g)$. Here, $\sigma_x$, $\sigma_z$ are Pauli spin matrices, $\epsilon$ is the ``bias'' of the asymmetric double well, and $\Delta_0$ is the ``bare tunneling matrix element''. In the canonical basis, whose vectors represent the states of the system localized in the left and the right well, $H_\s$ has the representation
\begin{equation}
H_\s = \frac 12
\left[
\begin{array}{cc}
\epsilon & -\Delta_0 \\
-\Delta_0 & -\epsilon
\end{array}
\right].
\label{sb1}
\end{equation}
The diagonalization of $H_\s$ yields $H_\s\cong {\rm diag}(E_+,E_-)$, where $E_\pm = \pm\frac 12\sqrt{\epsilon^2+\Delta_0^2}$. The operator $v=\sigma_z\otimes \varphi(g)$ is represented in the basis diagonalizing $H_\s$ as \fer{michael*}, with $a=-b= -(\frac{\Delta_0^2}{\epsilon^2}+1)^{-1/2}$ and $c=\frac 12 (\frac{\epsilon^2}{\Delta_0^2}+1)^{-1/2}$.

\medskip

{\it Illustration.\ } Let us discuss the decoherence and thermalization properties in the case where $\s$ is initially given by a coherent superposition in the energy basis. For sake of explicitness we take
\begin{equation}
\rhobar_0=\textstyle \frac 12
\left[
\begin{array}{cc}
 1 & 1\\
 1 & 1
\end{array}
\right].
\label{illust}
\end{equation}
We obtain the following expressions for the dynamics of matrix elements,
for all $t\geq 0$:
\begin{eqnarray*}
[\rhobar_t]_{m,m} &=& \frac{\e^{-\beta E_m}}{Z_{\s,\beta}} +\frac{(-1)^m}{2} \tanh\left(\frac{\beta\Delta}{2}\right) \e^{\i t\varepsilon_0(\lambda)} +R_{m,m}(\lambda,t) ,\ \ \ m=1,2, \\
{} [\rhobar_t]_{1,2} &=& {\textstyle \frac 12} \e^{\i t\varepsilon_{-\Delta}(\lambda)} +R_{1,2}(\lambda,t),
\end{eqnarray*}
where the numbers $\varepsilon$ are given in \fer{i4}. The remainder terms satisfy $|R_{m,n}(\lambda,t)|\leq C\lambda^2$, uniformly in $t\geq 0$, and they can be decomposed into a sum of a constant and a decaying part,
$$
R_{m,n}(\lambda,t) = \av{\av{p_{n,m}}}_\infty-\delta_{m,n} \frac{\e^{-\beta E_m}}{Z_{\s,\beta}} +R'_{m,n}(\lambda,t),
$$
where $|R'_{m,n}(\lambda,t)|=O(\lambda^2\e^{-\gamma t})$, with $\gamma=\min\{{\rm Im}\varepsilon_0, {\rm Im}\varepsilon_{\pm\Delta}\}$. These relations show in particular that, to second order in $\lambda$, convergence of the populations to the
equilibrium values (Gibbs law), and decoherence occur exponentially
fast, with rates $\tau_T=[{\mathrm Im}\varepsilon_0(\lambda)]^{-1}$
and $\tau_D=[{\mathrm Im}\varepsilon_\Delta(\lambda)]^{-1}$,
respectively. (If either of these imaginary parts vanishes then the
corresponding process does not take place, of course.) In
particular, coherence of the initial state stays preserved on time
scales of the order $\lambda^{-2}[|c|^2
\xi(\Delta)+(b-a)^2\xi(0)]^{-1}$, c.f. \fer{i4}. We show how to
arrive at the above expressions at the end of Section
\ref{subsqubit}.

\medskip

The method we use in this work yields an error estimate in \fer{introdd1} which is not uniform in $T=1/\beta\rightarrow 0$. However, our result can be upgraded to a uniform estimate by employing spectral renormalization group methods as developed in \cite{BFS,MMS1,MMS2,Mthesis}. This will be addressed elsewhere.

As mentioned above, we prove equation \fer{introdd1} using quantum statistical resonance theory, which is based on spectral deformation techniques. To keep the exposition as simple as possible we use in this paper the simplest and most restrictive version of this method, namely, the translation deformation. This produces weaker results (like non-uniformity in temperature of the error estimate in \fer{introdd1}, as mentioned above) than can be obtained by more refined techniques. Dilation deformation \cite{BFS} and a combination of dilation and translation deformation \cite{MMS1,MMS2} in conjunction with spectral renormalization group methods weakens the restrictions on the class of treatable interactions considerably and strengthens the results, but at the price of a much more involved mathematical machinery.

Although there is a subtle mathematical theory behind our
techniques, we believe that on a formal level, they are simpler and
more powerful than the standard path-integral methods
(\cite{CL,A,LO1,LO2}). The rigorous treatment based on master
equations and Lindblad generators \cite{LS} seems to be more
difficult. While the path-integral and master equation approaches
are intrinsically time-dependent, the resonance theory is formulated
as a stationary eigenvalue problem for some (albeit non-self-adjoint)
operators $\Lambda_{e,\beta,\lambda}$ acting on subspaces of two
copies of the system Hilbert space, $\hh_\s\otimes\hh_\s$. In
particular, the complex numbers $\varepsilon$ are eigenvalues of
these operators. See the remark at the end of Section
\ref{dynrestheory}.

This paper is organized as follows. In Section \ref{opensystsection}
we introduce the model and in Section \ref{intro1} we present our
main result for general $N-$level systems and for the qubit. We
develop a general dynamical resonance theory for systems at positive
temperatures and densities in Sections
\ref{dynrestheory}-\ref{rrrsect} (our general theorem stating
\fer{introdd1} is proven in Section \ref{dynrestheory} and the
relation \fer{int1} is shown in Section \ref{sect}). We give two
applications of this theory in Section \ref{sect4}: the first one
yields a proof of the results for the qubit mentioned above, the
second one illustrates our method on an explicitly solvable
non-demolition model. Appendices \ref{appgluing} and \ref{appA} contain explicit formulas for some
quantities which are important for our resonance theory. In Appendices \ref{AppCor} and \ref{prooflemmalso} we outline the perturbation theory of equilibrium states, and we give
the proofs of several propositions of previous sections, including
the short calculation of the explicit solution of a non-demolition
model.

\section{Open quantum systems}
\label{opensystsection}

{\bf The system $\s$.\ } Let $\hh_\s$ be the space of pure states of a quantum system $\s$ whose dynamics is generated by a Hamiltonian $H_\s$. In applications we will take $\hh_\s=\cx^N$ and $H_\s={\rm diag}(E_1,\ldots,E_N)$, but we give a discussion of more general systems $\s$ with finitely or infinitely many (discrete) levels. This discussion  is straightforward and instructive, we believe. The evolution of a density matrix $\rho_\s$ on $\hh_\s$ is
\begin{equation}
t \mapsto  \e^{-\i tH_\s}\rho_\s \e^{\i tH_\s},\ \ t\in\rx.
\label{dyns}
\end{equation}
The Gibbs state at temperature $T=1/\beta>0$ is $\rho_{\s,\beta}=Z_\beta^{-1}\e^{-\beta H_\s}$, with the normalization constant $Z_\beta=\tr_\s\e^{-\beta H_\s}$. (If $\dim\hh_\s<\infty$ then $Z_\beta<\infty$ for any $H_\s$. In the infinite-dimensional case we {\it assume} that $H_\s$ is trace-class.) Observables of $\s$ are operators on $\hh_\s$, they form the algebra of all bounded operators ${\cal B}(\hh_\s)$.

One can represent any state on $\s$, mixed or pure, by a single vector in the Hilbert space $\h_\s=\hh_\s\otimes\hh_\s$. This is the so-called Gelfand--Naimark--Segal representation of states. To see how this works take an arbitrary density matrix $\rho_\s$ on $\hh_\s$, and write it in its diagonalizing basis as
\begin{equation}
\rho_\s = \sum_{j} p_j |\psi_j\rangle\langle\psi_j|.
\label{densmat}
\end{equation}
The corresponding state is defined by $A\mapsto \tr_\s(\rho_\s A)$ for any observable $A\in {\cal B}(\hh_\s)$. Let us denote by $L^2(\hh_\s)$ the set of all Hilbert--Schmidt operators, i.e., $A\in L^2(\hh_\s)$ if and only if $\tr_\s(A^*A)$ is finite. (In the case $\dim\hh_\s<\infty$  we have $L^2(\hh_\s)={\cal B}(\hh_\s)$.) $L^2(\hh_\s)$ is a Hilbert space with the scalar product $\scalprod{A}{B}_2=\tr_\s(A^*B)$. Since the density matrix $\rho_\s$ is a positive trace class operator, its square is a Hilbert--Schmidt operator, and due to the cyclicity of the trace we have
\begin{equation}
\tr_\s(\rho_\s A) = \scalprod{\sqrt{\rho_\s}}{A\sqrt{\rho_\s}}_2.
\label{**}
\end{equation}
This shows that the density matrix $\rho_\s$ on $\hh_\s$ is represented by the vector $\sqrt{\rho_\s} \in L^2(\hh_\s)$. Instead of working with the state space $L^2(\hh_\s)$ it is often convenient to switch to $\h_\s=\hh_\s\otimes\hh_\s$. This is done via the correspondence
\begin{equation}
T: |\varphi\rangle\langle\psi| \longrightarrow \varphi\otimes\cc\psi,
\label{correspondence}
\end{equation}
which extends by linearity to an isometric isomorphism between $L^2(\hh_\s)$ and $\h_\s$. Here, $\cc$ is a map on $\hh_\s$ that is chosen to be antilinear (conjugate linear) since $T$ should be linear but  $\psi\mapsto \langle\psi|$ is antilinear. Furthermore, to make $T$ an isometry (norm preserving) $\cc$ has to be antiunitary, meaning that it satisfies
\begin{equation}
\scalprod{\cc\chi_1}{\cc\chi_2}=\overline{\scalprod{\chi_1}{\chi_2}}
\label{au}
\end{equation}
for all $\chi_{1,2}\in\hh_\s$. Relation \fer{au} implies that $\cc$ is bijective: injectivity follows from $\|\cc \chi\|=\|\chi\|$ and surjectivity follows from the fact that $\cc^2$ is unitary, so ${\mathrm Ran}\cc\supseteq{\mathrm Ran}\cc^2=\hh_\s$.

In what follows we choose $\cc$ to be the operator that takes the complex conjugate of coordinates w.r.t. the basis of $\hh_\s$ in which $H_\s$ is diagonal. In this case $\cc$ is an involution, $\cc^2=\cc$.

Since $T$ is an isomerty we have $\scalprod{A}{A}_2 = \|T A\|^2=\scalprod{TA}{TA}$, where the norm and the scalar product on the r.h.s. are those of $\h_\s$. This implies that $\scalprod{A}{B}_2=\scalprod{TA}{TB}$, by the polarization identity. In particular, \fer{**} gives
\begin{equation}
\tr_\s(\rho_\s A) = \scalprod{T\sqrt{\rho_\s}}{T A\sqrt{\rho_\s}}.
\label{65}
\end{equation}
The vector in $\h_\s$ representing the state \fer{densmat} is thus
\begin{equation}
\Psi = T\sqrt{\rho_\s} = \sum_j \sqrt{p_j}\  \psi_j\otimes\cc\psi_j,
\label{61}
\end{equation}
and the Gibbs state corresponds to
\begin{equation}
\Omega_{\s,\beta} = Z_\beta^{-1/2}\sum_j \e^{-\beta E_j/2} \varphi_j\otimes\varphi_j,
\label{62}
\end{equation}
where $\varphi_j$ is the eigenvector of $H_\s$ corresponding to the energy $E_j$.

Due to \fer{65} and since $TA\sqrt{\rho_\s} = (A\otimes\bbbone)\Psi$ we have
\begin{equation}
\tr_\s(\rho_\s A) = \scalprod{\Psi}{(A\otimes\bbbone) \Psi}.
\label{64}
\end{equation}
The algebra of observables of $\s$ is given by ${\cal B}(\hh_\s)\otimes \bbbone$ when viewed as operators on $\h_\s$.

{\it Remark.\ }
It appears that in the representation of the Gibbs state the temperature dependence (parameter $\beta$) is entirely concentrated on the vector $\Omega_{\s,\beta}$, \fer{62}, and the represented observables $A\otimes\bbbone$ are independent of $\beta$. We may transfer the $\beta$-dependence entirely (or partly) to the observables by a change of basis. For example, let $\Psi'$ be any fixed vector on $\h_\s$ and take any unitary $U_\beta$ with the property $U_\beta\Omega_{\s,\beta}=\Psi'$. Then we have $\tr_\s(\rho_\s A) = \scalprod{\Psi'}{A_\beta\,\Psi'}$, where $A_\beta= U_\beta(A\otimes\bbbone)U^*_\beta$.

\medskip

As we see from \fer{dyns}, \fer{densmat} and \fer{61}, the evolution of $\Psi$ is given by
\begin{equation}
t\mapsto \sum_j \sqrt{p_j}\  \e^{-\i tH_\s}\psi_j\otimes \cc\e^{-\i tH_\s}\psi_j = \e^{-\i tL_\s}\Psi,
\label{60}
\end{equation}
where
\begin{equation}
L_\s = H_\s\otimes\bbbone - \bbbone\otimes H_\s
\label{63}
\end{equation}
is called the (standard) Liouville operator. It satisfies the relation
\begin{equation}
L_\s\Omega_{\s,\beta}=0.
\label{63.1}
\end{equation}

In the Heisenberg picture, a system observable $A$ evolves according to
\begin{equation}
t\mapsto \e^{\i tL_\s} (A\otimes\bbbone) \e^{-\i tL_\s}.
\label{heisenberg}
\end{equation}

{\it Remark on the choice of the Liouville operator.\ } The map $T$, \fer{correspondence}, is only a particular choice of an isometric isomorphism. We may more generally put
$$
T: |\varphi\rangle\langle\psi| \longrightarrow U\varphi\otimes V\cc\psi,
$$
where $U$ and $V$ are arbitrary unitary operators on $\hh_\s$. We may even define a {\it time-dependent} isometric isomorphism $T_t$, by taking time-dependent families of unitaries $U_t$ and $V_t$. For instance, the special choice $U=\bbbone$ and $V=V_t$ yields
\begin{equation*}
T_t A\e^{-\i tH_\s}\sqrt{\rho_\s}\e^{\i tH_\s} = (A\otimes\bbbone)(\e^{-\i tH_\s}\otimes V_t \e^{\i tH_\s}) \Psi,
\end{equation*}
with $\Psi$ given in \fer{61}. Therefore, we may equally well define the evolution of $\Psi$ by $t\mapsto \Psi(t)=(\e^{-\i t H_\s}\otimes V_t \e^{\i tH_\s})\Psi$. As an example, take $V_t=\e^{-\i tH_\s}$, then $\Psi(t)=(\e^{-\i t H_\s}\otimes \bbbone) \Psi$. It is easy to see that the only choice of $V_t$ for which the Gibbs vector $\Omega_{\s,\beta}$, \fer{62}, is invariant under the evolution, i.e. for which $\Omega_{\s,\beta}(t)=\Omega_{\s,\beta}$ for all $t$, is given by $V_t=\bbbone$. This convenient choice results in \fer{60}, \fer{63} and \fer{63.1}.

\bigskip

{\bf The reservoir $\r$.\ } We introduce  now a second quantum system, a reservoir $\r$ in a state of thermal equilibrium at a temperature $T=1/\beta>0$. A typical reservoir is a very large quantum system, say a quantum gas with a given particle density (or a radiation field). In order to suppress recurrences one may consider the idealized situation of an infinitely extended reservoir giving rise to truly irreversible processes (like thermalization or decoherence). In physical experiments reservoirs have of course finite size, but the idealized limit is a good approximation for physically realistic times that are not exceedingly large, see e.g. \cite{FMU} for a discussion of this point.

The reservoir $\r$ we consider is the infinitely extended gas of massless relativistic Bosons (of photons or phonons, for example) at positive temperature and positive density. Our approach applies also to fermionic reservoirs, and in fact becomes technically much simpler in that case. The state of $\r$ is obtained by performing a thermodynamic limit of finite-volume equilibrium (Gibbs) states with fixed temperature and density. We refer to \cite{Miqg} for a detailed exposition of this matter in textbook-style. Just as in the case of the system $\s$ one can represent the equilibrium state by a single vector in a suitable Hilbert space. This is the so-called Araki--Woods representation, \cite{AW, Miqg}, in which the Hilbert space is given by
\begin{equation}
\h_\r = {\cal F}(L^2(\rx^3,\d^3k)) \otimes {\cal F}(L^2(\rx^3,\d^3k)),
\label{66}
\end{equation}
where ${\cal F}(L^2(\rx^3,\d^3k))$ is the bosonic Fock space over the one-particle space of wave functions $L^2(\rx^3,\d^3k)$ in momentum representation. The usual bosonic creation and annihilation operators $a^*(k)$ and $a(k)$, $k\in\rx^3$, are represented in $\h_\r$ as the thermal creation- and annihilation operators,
\begin{eqnarray}
a(k)\mapsto a_\beta(k) &=& \sqrt{1+\mu_\beta} \ a(k)\otimes \bbbone + \sqrt{\mu_\beta}\ \bbbone\otimes a^*(k),\label{corr1}\\
a^*(k)\mapsto a^*_\beta(k) &=& \sqrt{1+\mu_\beta} \ a^*(k)\otimes \bbbone + \sqrt{\mu_\beta}\ \bbbone\otimes a(k),\label{corr2}
\end{eqnarray}
where $\mu_\beta$ is Planck's momentum density distribution for black-body radiation,
\begin{equation}
\mu_\beta(k) =\frac{1}{\e^{\beta |k|}-1}
\label{planck}
\end{equation}
(we take the Bose gas in a phase without a Bose--Einstein condensate). The representation \fer{corr1}, \fer{corr2} is the equivalent to the representation of an observable $A\in{\cal B}(\hh_\s)$ by $A\otimes\bbbone\in {\cal B}(\h_\s)$ in the case of the system $\s$, c.f. the remark after \fer{64}. The map defined by \fer{corr1}, \fer{corr2} is called the Araki-Woods representation map. 

Denote by $\Omega$ the vacuum vector of ${\cal F}(L^2(\rx^3,\d^3k))$. It is easily seen that the vector
\begin{equation}
\Omega_{\r,\beta} = \Omega\otimes\Omega
\label{67}
\end{equation}
represents the equilibrium state and 
$$
\scalprod{\Omega_{\r,\beta}}{a_\beta^*(k)a_\beta(l)\Omega_{\r,\beta}} =\delta(k-l)\mu_\beta(k),
$$
where $\delta$ is the Dirac delta function.

The dynamics of a density matrix $\rho_\r$ of the reservoir (acting on $\h_\r$) is given by
\begin{equation}
t \mapsto  \e^{-\i tL_\r} \rho_\r \e^{\i tL_\r},
\label{68}
\end{equation}
where
\begin{equation}
L_\r = H_\r\otimes\bbbone - \bbbone\otimes H_\r,
\label{69}
\end{equation}
with
\begin{equation}
H_\r = \int_{\rx^3} |k| a^*(k)a(k)\d^3k.
\label{70}
\end{equation}

The self-adjoint operator $L_\r$ is called the standard Liouville operator for $\r$, it satisfies (compare with \fer{63.1})
\begin{equation}
L_\r\Omega_{\r,\beta}=0.
\label{71}
\end{equation}

The smoothed-out thermal creation and annihilation operators are defined as
\begin{eqnarray}
a^*_\beta(h)&=&\int_{\rx^3} h(k) a^*_\beta(k) \d^3k,\label{80}\\
a_\beta(h) &=& \int_{\rx^3} \overline h(k) a_\beta(k)\d^3k\label{81},
\end{eqnarray}
where $h\in L^2(\rx^3,\d k^3)$ is a wave function of a single Boson and where the $a_\beta(k)$ and $a_\beta^*(k)$ are given in \fer{corr1}, \fer{corr2}.

It is not hard to check explicitly that $L_\r$, \fer{69}, implements the {\it Heisenberg dynamics of observables}, given by the Bogoliubov transformation
$$
t\mapsto \e^{\i tL_\r} a_\beta^\#(h) \e^{-\i tL_\r} = a_\beta^\#(\e^{\i |k| t}h).
$$

\medskip

{\bf The total system $\s+\r$.\ } The joint system $\s + \r$ is described by the Hilbert space
\begin{equation}
\h=\h_\s\otimes\h_\r,
\label{73}
\end{equation}
and the non-interacting dynamics of a density matrix $\rho$ on $\h_\s\otimes\h_\r$ is
\begin{equation}
t\mapsto \e^{-\i tL_0} \rho \e^{\i tL_0},
\label{74}
\end{equation}
with
\begin{equation}
L_0 = L_\s + L_\r.
\label{75}
\end{equation}
The state $\rho_{\s,\beta}\otimes \rho_{\r,\beta}=|\Omega_{\s,\beta}\rangle\langle \Omega_{\s,\beta}|\otimes |\Omega_{\r,\beta}\rangle\langle \Omega_{\r,\beta}|$ is an equilibrium state w.r.t. the non-interacting dynamics.

The coupling between $\s$ and $\r$ is specified by an interaction operator $V$ which is an observable of the joint system and is related to the interaction operator $v$ given in \fer{michael*}. The full dynamics of a density matrix $\rho$ on $\h_\s\otimes\h_\r$,
\begin{equation}
t\mapsto \e^{-\i t L_\lambda} \rho \e^{\i t L_\lambda},
\label{72}
\end{equation}
is generated by
\begin{equation}
L_\lambda = L_0 +\lambda V,
\label{73.1}
\end{equation}
where $\lambda$ is a coupling constant. An important class of interactions is given by coupling operators of the form 
\begin{equation}
v= G\otimes\varphi(g),
\label{2.28a}
\end{equation}
see also \fer{ii1}, where $G$ is a self-adjoint operator on $\hh_\s$ and where
\begin{equation}
\varphi(g) =\frac{1}{\sqrt 2}\left( a^*(g) +a(g)\right)
\label{fieldoperator}
\end{equation}
is the bosonic field operator on ${\cal F}(L^2(\rx^3,\d^3k))$, smeared out with a form factor $g\in L^2(\rx^3,\d^3k)$. On the positive temperature Hilbert space $\h_\s\otimes\h_\r$ this interaction Hamiltonian leads to the operator
\begin{equation}
V = G\otimes\bbbone_\s\otimes\varphi_\beta(g),
\label{31}
\end{equation}
where
\begin{equation}
\varphi_\beta(g)=\frac{1}{\sqrt 2}(a_\beta^*(g)+a_\beta(g))
\label{83}
\end{equation}
is the smoothed out thermal field operator, \fer{80}, \fer{81}. More generally, the interaction is specified by an operator $V\in\mm=\mm_\s\otimes\mm_\r$, or a $V$ which is unbounded, but affiliated  with $\mm$,\footnote{meaning that it commutes with all bounded operators belonging the the commutant algebra $\mm'$} like \fer{31}.

For a wide class of interactions  $V$,  the interacting dynamics admits an equilibrium state $\rho(\beta,\lambda)$ (at least for small $\lambda$), but we do not limit our discussion to such operators $V$.

The {\it reduced density matrix} $\rhobar_t$ of the open system $\s$ is obtained by tracing out the degrees of freedom of the reservoir, it is defined by
$$
\tr_{\s}(\rhobar_t A) = \tr_{\s+\r}\left( \e^{-\i t L_\lambda} \rho_0 \e^{\i tL_\lambda}(A\otimes\bbbone_\r) \right),
$$
for all $A\in {\cal B}(\hh_\s)$, where $\rho_0$ is the initial state of $\s$. It is customary to take $\rho_0=\rhobar_0\otimes\rho_{\r,\beta}$, where $\rhobar_0$ is the initial state of $\s$ and  $\rho_{\r,\beta}$ is the reservoir equilibrium state, but our analysis works equally well for initially entangled (non-product) states which are arbitrary local perturbations of $\rho_{\r,\beta}$.

Let $\{\varphi_j\}_{j\geq 1}$  be an orthonormal basis of $\hh_\s$ diagonalizing $H_\s$. The reduced density matrix has matrix elements
$$
[\rhobar_t]_{m,n} := \scalprod{\varphi_m}{\rhobar_t\varphi_n}=\tr_\s(\rhobar_t p_{n,m})
$$
where
\begin{equation}
p_{n,m}=|\varphi_n\rangle\langle\varphi_m|.
\label{pmn}
\end{equation}

We say that the system $\s+\r$ has the property of {\it return to equilibrium} iff
$$
\lim_{t\rightarrow\infty}\tr_{\s+\r}(\rho_0 \e^{\i t L_\lambda}  M \e^{-\i tL_\lambda}) = \tr_{\s+\r}(\rho(\beta,\lambda) M),
$$
for all observables (of the joint system) $M$ and for all initial density matrices $\rho_0$ on $\h$. The large time limit of the reduced density matrix of such a system is given by
$$
\rhobar_\infty:=\lim_{t\rightarrow\infty}\rhobar_t = \tr_\r(\rho(\beta,\lambda)) = \rho_{\s,\beta} + O(\lambda),
$$
since, by perturbation theory of equilibrium states, $\rho(\beta,\lambda) = \rho_{\s,\beta}\otimes\rho_{\r,\beta} +O(\lambda)$. The leading term of $\rhobar_\infty$ (for small coupling) is just the Gibbs state of $\s$. In this sense, $\s$ undergoes the process of {\it thermalization}.

The system $\s$ is said to exhibit (full) {\it decoherence (in the energy basis)} if the off-diagonal matrix elements of the reduced density matrix vanish in the limit of large times,
$$
\lim_{t\rightarrow\infty} [\rhobar_t]_{m,n} = 0
$$
whenever $m\neq n$.

It has to be pointed out that $\rhobar_\infty$ is not diagonal in the energy basis in general, but only its leading part is. Therefore, thermalization prevents full decoherence since the final state of the thermalized system $\s$ has still non-vanishing off-diagonal elements, which are of the order $O(\lambda)$. In fact, there is {\it creation of non-vanishing off-diagonals by thermalization} since even if $\s$ is initially an incoherent superposition of basis elements of a given basis $\{\chi_j\}_{j\geq 1}$,
$$
\rhobar_0 = \sum_m \alpha_m|\chi_m\rangle\langle\chi_m|,
$$
it will become a coherent superposition since in general,
$$
\scalprod{\chi_m}{\rhobar_\infty\chi_n} \neq 0.
$$
We illustrate this for a qubit in Theorem \ref{cor1}.

Full decoherence can occur if thermalization processes are excluded. For instance, if one suppresses energy exchanges between $\s$ and $\r$ (i.e., if $V$ commutes with $H_\s$) then, even if an equilibrium state exists, the system does not converge to it as $t\rightarrow\infty$. So there is no thermalization but full decoherence, i.e. $[\rhobar_\infty]_{m,n}=0$, $m\neq n$, can be shown to occur in various models. (That there is no thermalization is readily seen since the populations are time-independent).

However, it is well known \cite{PSE,MP} that even if energy exchange processes are suppressed, full decoherence may not take place if the infrared modes of the reservoir are only weakly coupled to the system, see Section \ref{sectapp2}.

\section{Decoherence and thermalization of a general $N$-level system and of a qubit}
\label{intro1}

We analyze an $N$-level system coupled to a thermal environment modeled by an infinitely extended free massless relativistic Bose field at temperature $T>0$. Under suitable conditions on the interaction the system has the property of return to equilibrium. Our goal is to compare the thermalization and decoherence processes and, in particular, the speed of convergence of the diagonal and off-diagonal matrix elements of the reduced density matrix.

The state space of an $N$-level system is given by $\hh_\s=\cx^N$, and its Hamiltonian is
\begin{equation}
H_\s = {\rm diag}(E_1,\ldots,E_N).
\label{r-1}
\end{equation}
We couple the $N$-level system to the reservoir through an operator $v=G\otimes\varphi(g)$, where $G$ is a hermitian $N\times N$ matrix, and the field operator is as in \fer{fieldoperator}, with a coupling function $g(k)$, $k\in\rx^3$. Let $g(k)=g(r,\sigma)$, where $(r,\sigma)\in \rx_+\times S^2$. We make a regularity assumption on $g$. Fix any phase $\phi\in\rx$ and define
\begin{equation}
g_\beta(u,\sigma) := \sqrt{\frac{u}{1-\e^{-\beta u}}}\ |u|^{1/2}
\left\{
\begin{array}{ll}
g(u,\sigma) & \mbox{if $u\geq 0$},\\
-\e^{\i\phi} \overline g(-u,\sigma) & \mbox{if $u<0$},
\end{array}
\right.
\label{7}
\end{equation}
where $u\in\rx$ and $\sigma \in S^2$. The phase $\phi$ is a parameter which can be chosen appropriately to satisfy the following condition for a given coupling function $g$, see \cite{FM1}.

\medskip
{\bf (A)} We assume that the map $\omega\mapsto g_\beta(u+\omega,\sigma)$ has an analytic extension to a complex neighbourhood $\{|z|<\omega'\}$ of the origin, as a map from $\cx$ to $L^2(\rx^3,\d^3k)$.

\medskip
This condition ensures that the simplest version of a dynamical resonance theory - the one using complex deformations - can be implemented in a straightforward way. Examples of $g$ satisfying (A) are given by  $g(r,\sigma) = r^p \e^{-r^m} g_1(\sigma)$, where $p=-1/2+n$, $n=0,1,\ldots$, $m=1,2$, and $g_1(\sigma)=\e^{\i\phi}\overline g_1(\sigma)$.

The technical simplicity of the complex translation method comes at a price. On the one hand, it limits the class of admissible functions $g(k)$, which have to behave appropriately in the infra-red regime so that the parts of \fer{7} fit nicely together at $u=0$, to allow for an analytic continuation. On the other hand, the square root in \fer{7} must be analytic as well, which implies the condition $\omega'<2\pi/\beta$.

We now state our main result on the dynamics. Let $\psi_0\otimes\Omega_{\r,\beta}$ be the vector in $\h_\s\otimes\h_\r$ representing the density matrix at time $t=0$. Let $B$ be the unique operator in $\bbbone_\s\otimes{\cal B}(\cx^N)$ satisfying 
$$
B\Omega_{\s,\beta}=\psi_0.
$$
The existence of such a $B$ is not hard to verify, see also the end of Section \ref{subsqubit} for concrete examples. Define the vector 
$$
\Omega_{\beta,0} := \Omega_{\s,\beta}\otimes\Omega_{\r,\beta}.
$$ 
\begin{theorem}[Dynamical resonance theory]
\label{prop3.2}
Assume condition (A) with a fixed $\omega'$ satisfying $0<\omega'<2\pi/\beta$. There is a constant $c_0$ s.t. if $|\lambda|\leq c_0/\beta$ then the limit $\av{\av{A}}_\infty$, \fer{ergav}, exists for all observables $A\in{\cal B}(\hh_\s)$. Moreover, for all such $A$ and for all $t\geq 0$ we have 
\begin{eqnarray}
\lefteqn{
\av{A}_t - \av{\av{A}}_\infty
=\sum_{s\, :\  \varepsilon_0^{(s)}\neq 0}\e^{\i t\varepsilon_0^{(s)}} \scalprod{(B^*\psi_0)\otimes \Omega_{\r,\beta}}{Q_0^{(s)} (A\otimes\bbbone_\s) \Omega_{\beta,0}} }\nonumber\\
&&+\sum_{e\neq 0} \sum_{s=1}^{\nu(e)}\e^{\i t\varepsilon_e^{(s)}} \scalprod{(B^*\psi_0)\otimes \Omega_{\r,\beta}}{Q_e^{(s)} (A\otimes\bbbone_\s) \Omega_{\beta,0}}\nonumber \\
&& +O(\lambda^2\e^{-\frac t2[\max_{e,s}\{\rm Im\}\,\varepsilon_e^{(s)}+\omega'/2]}).
\label{24}
\end{eqnarray}
Here, the $\varepsilon_e^{(s)}$ are given by \fer{int1}, $1\leq \nu(e) \leq {\rm mult}(e)$ counts the splitting of the eigenvalue $e$ into distinct resonance energies $\varepsilon_e^{(s)}$, and the $Q_e^{(s)}$ are (non-orthogonal) finite-rank projections. 
\end{theorem}

{\it Remarks.\ }
1. One can explicitly expand in powers of $\lambda$ both the resonance energies $\varepsilon_e^{(s)}$ and the resonance projections $Q_e^{(s)}$, see Sections \ref{sect} and \ref{rrrsect}.

2. Our techniques are not restricted to unentangled initial states (see also \fer{i11}, \fer{i22}).

Return to equilibrium for $N$-level systems with a coupling \fer{31} to the Bose field has been shown by several authors over the last few years, provided the coupling is sufficiently effective,  and under regularity conditions much weaker than (A), see for instance \cite{JP1,BFS,FM1} and references therein. For the sake of notational simplicity we will restrict our attention to the qubit ($N=2$) in the presentation of the remaining results in this section. The Hamiltonian of the qubit is $H_\s={\rm diag}(E_1,E_2)$, we set $\Delta:=E_2-E_1>0$. The operator $G$ in \fer{31} is represented in the energy-basis by the matrix
\begin{equation}
G =
\left[
\begin{array}{cc}
a & c \\
\overline c & b
\end{array}
\right],
\label{32}
\end{equation}
where $a,b\in\rx$. 

\begin{theorem}[Thermalization, return to equilibrium for the qubit]
\label{thmrte}
Assume condition (A), and that $|c|^2\xi(\Delta)>0$. There is a constant $c_0$ s.t. if $0<|\lambda|<c_0/\beta$  then  the qubit coupled to the reservoir of thermal Bosons has the property of return to equilibrium.
\end{theorem}

The condition $|c|^2\xi(\Delta)>0$ is often called the ``Fermi Golden Rule Condition''. It means that the interaction induces second-order ($\lambda^2$) energy exchanging processes at the Bohr frequency of the qubit (emission and absorption of reservoir quanta). The condition $c\neq 0$ is actually {\it necessary} for thermalization while $\xi(\Delta)>0$ is not (higher order processes can drive the system to equilibrium).

\begin{theorem}[Creation of off-diagonals by thermalization]
\label{cor1}
Under the assumption of Theorem \ref{thmrte} the off-diagonal elements of the reduced density matrix are given in the limit $t\rightarrow\infty$, irrespectively of the initial density matrix $\rho_0$, by
\begin{eqnarray*}
[\rhobar_\infty]_{1,2} = \frac{c\lambda^2}{Z_{\s,\beta}}\scalprod{g}{(a \e^{-\beta E_1} B_1 + b \e^{-\beta E_2} B_2)g} +O(\lambda^4),
\end{eqnarray*}
where $Z_{\s,\beta}=\tr_\s(\e^{-\beta H_\s})$ is the partition function of $\s$, and where 
\begin{eqnarray*}
B_1&=& \frac{\mu}{\omega(\omega+\Delta)}\Big[ 
\e^{\beta\omega/2} (\e^{\beta\omega/2}-1) - \e^{-\beta\Delta/2}(\e^{-\beta\Delta/2}-1) +\e^{\beta\Delta/2}-1 \Big]\\
&&+\ \frac{1+\mu}{\omega(\omega-\Delta)} (\e^{-\beta\omega/2} -\e^{-\beta\Delta/2})(\e^{-\beta\Delta/2}+\e^{-\beta\omega/2}-1)\\
&&- \ \frac{\e^{-\beta\Delta/2}-1}{\omega\Delta} (\e^{-\beta\Delta/2}-\mu-1)
+ \frac{\mu}{\Delta(\omega+\Delta)}(\e^{-\beta\Delta/2}-1),
\end{eqnarray*}
and
\begin{eqnarray*}
B_2&=& \frac{1+\mu}{\omega(\omega+\Delta)}\Big[ (\e^{-\beta\omega/2}-\e^{\beta\Delta/2})^2 +\e^{\beta\Delta/2}-1\Big]\\
&& +\frac{\mu}{\omega(\omega-\Delta)}\Big[ \e^{\beta\omega} -\e^{\beta\Delta/2}(\e^{\beta\Delta/2}-1) -1\Big] +   \frac{\e^{\beta\Delta/2} (\e^{\beta\Delta/2}-1)}{\omega\Delta}\\
&& +\  \frac{1+\mu}{\Delta(\omega+\Delta)}(\e^{\beta\Delta/2}-1) 
- \frac{\mu}{\Delta(\omega-\Delta)} (\e^{\beta \Delta/2}-1).  
\end{eqnarray*}
Here $\omega=|k|$, $\Delta=E_2-E_1>0$ is the energy gap of $H_\s$, and $\mu=\mu_\beta$ is given in \fer{planck}.
\end{theorem}
This theorem follows from an expansion of the equilibrium state in powers of $\lambda$. We give an outline of the proof in Appendix \ref{AppCor}.

{\it Remarks.\ }
1)\ Even if the system $\s$ starts out in an incoherent superposition of vectors from the energy basis, Theorem \ref{cor1} shows that in the large-time limit, coherence of order $\lambda^2$ is built up in that basis.

2) \ If the interaction matrix $G$ is diagonal, i.e. if $c=0$, then the $O(\lambda^2)$ terms in the off-diagonals of $\rhobar_\infty$ vanish. The same happens if $a=b=0$, i.e., if $G$ is purely off-diagonal.

Recall the notation  $\av{A}_t$, $\av{\av{A}}_\infty$ and $p_{n,m}$, \fer{avA}, \fer{ergav} and \fer{pmn}.

\begin{theorem}[Convergence of matrix elements]
\label{thm1}
Assume (A) for a fixed $\omega'>0$ satisfying $0<\omega'<2\pi/\beta$.
There is a constant $c_0$ s.t. the following statements hold for $|\lambda|<c_0/\beta$. For any initial density matrix of the form $\rhobar_0\otimes\rho_{\r,\beta}$ and any observable $A$ of $\s$ the limit $\av{\av{A}}_\infty$ exists, and statements \fer{i1}-\fer{i4} about the asymptotic behaviour of $[\rhobar_t]_{m,n}$ are true.
\end{theorem}
Let us again point out that our method works for any entagled initial density matrix $\rho_0$ on $\h_\s\otimes\h_\r$, c.f. \fer{i11}, \fer{i22}.

\section{Dynamical resonance theory for systems at positive temperatures and positive densities and proof of Theorem \ref{prop3.2}}
\label{dynrestheory}

The goal of this section is to arrive at relation \fer{24} expressing the asymptotic behaviour of the averages of a system observable $A$ as $t\rightarrow\infty$.

The state of $\s$, given by the initial density matrix $\rhobar_0$, is represented by the vector state $A\mapsto \scalprod{\psi_0}{(A\otimes\bbbone_\s) \,\psi_0}$, for a $\psi_0\in\h_\s=\hh_\s\otimes\hh_\s$, see \fer{61}. For instance, $\psi_0=\Omega_{\s,\beta}$ if $\s$ is initially in equilibrium, c.f. \fer{62}.

Let the initial state of the total system $\s+\r$ be the product state $\rhobar_0\otimes\rho_{\r,\beta}$, which is represented on $\h_\s\otimes\h_\r$ by the vector
\begin{equation}
\psi_0\otimes\Omega_{\r,\beta}\in \h_\s\otimes\h_\r,
\label{productic}
\end{equation}
where $\Omega_{\r,\beta}$ is given in \fer{67}. An observable $A\in\mm$ of the system $\s+\r$ evolves according to the Heisenberg evolution
$$
t\mapsto \e^{\i tL_\lambda} A\e^{-\i tL_\lambda},
$$
where $L_\lambda$ is given in \fer{73.1}, \fer{31}. The Schr\"odinger dynamics of a vector $\psi\in\h_\s\otimes\h_\r$ determining the state $\scalprod{\psi}{\ \cdot\ \psi}$ on $\mm$ is given by $t\mapsto \psi_t=\e^{-\i tL_\lambda}\psi$. The average of an observable $A$ of $\s$ at time $t$ is thus
\begin{equation}
\av{A}_t = \scalprod{\psi_0\otimes\Omega_{\r,\beta}}{\e^{\i tL_\lambda} (A\otimes\bbbone_\s\otimes\bbbone_\r)\e^{-\i tL_\lambda}\, \psi_0\otimes\Omega_{\r,\beta}}.
\label{97}
\end{equation}
It is reasonable to just write $A$ instead of $A\otimes\bbbone_\s\otimes\bbbone_\r$ in this section.
Our goal is to examine  $\av{A}_t$ as a function of $t$.

It is not hard to see that $\Omega_{\s,\beta}$ has the following property of separability: given any vector $\psi_0\in\h_\s\otimes\h_\s$ we can find a unique operator $B\in\bbbone_\s\otimes{\cal B}(\hh_\s)$ with the property that
\begin{equation}
\psi_0 = B\Omega_{\s,\beta}.
\label{96}
\end{equation}
One can solve this equation for the matrix elements of $B$, e.g. by using \fer{62}. See also the end of Section \ref{subsqubit} for concrete examples.

An entangled initial state of the system, given by a density matrix $\rho_0$, is realized on the Hilbert space $\h_\s\otimes\h_\r$ by a vector $\Psi_0$ which is not of the product \fer{productic}. However, one can show that any $\Psi$ can be approximated arbitrarily well by vectors of the form $B\Omega_{\beta,0}$, for some $B$ in the commutant of $\mm$, and where
\begin{equation}
\Omega_{\beta,0} = \Omega_{\s,\beta}\otimes\Omega_{\r,\beta}.
\label{3.1}
\end{equation}
(The commutant of $\mm$ is the von Neumann algebra consisting of all operators on $\h_\s\otimes\h_\r$ which commute with all operators of $\mm$.) The following arguments can then be carried out in a similar fashion. This is why our method works for arbitrary initial states, not only for Feynman-Vernon type, disentangled initial states.

Since $\e^{\i tL_\lambda}\mm\e^{-\i tL_\lambda}=\mm$ (invariance of the algebra of observables), the operator $B$ (belonging to the commutant of $\mm$,) commutes with $\e^{\i tL_\lambda} (A\otimes\bbbone_\s\otimes\bbbone_\r)\e^{-\i tL_\lambda}$. We can thus use \fer{96} in \fer{97} and commute $B$ to the left, with the result
\begin{equation}
\av{A}_t = \scalprod{(B^*\psi_0)\otimes\Omega_{\r,\beta}}{\e^{\i tL_\lambda} A \e^{-\i tL_\lambda} \Omega_{\beta,0}}.
\label{3}
\end{equation}

The following formula, derived in \cite{MMS1,MMS2}, reduces the description of the long-time behaviour of \fer{3} to a spectral problem for an auxiliary operator $K_\lambda(\omega)$ defined below,
\begin{equation}
\av{A}_t = \frac{-1}{2\pi\i} \int_{\rx-\i}  \e^{\i tz} \scalprod{(B^*\psi_0)\otimes\Omega_{\r,\beta}}{(K_\lambda(\omega)-z)^{-1} A \Omega_{\beta,0}} \d z.
\label{16}
\end{equation}
The integral in \fer{16} is understood as an improper Riemann integral,
\begin{equation}
\int_{\rx -\i}f(z) \d z = \lim_{a,b\rightarrow\infty}\int_{-a}^{b} f(x-\i) \d x.
\label{iri}
\end{equation}
We present here a heuristic derivation of \fer{16}, see however \cite{MMS1,MMS2} for a rigorous proof.
First we get rid of the factor $\e^{-\i tL_\lambda}$ in \fer{3} by the following trick \cite{JP2,MMS1,MMS2}: one can add to $L_\lambda$ a term which does not change the dynamics but which is s.t. the resulting sum $K_\lambda$ satisfies $K_\lambda\Omega_{\beta,0}=0$. The operator $K_\lambda$ is of the form 
$$
K_\lambda = L_0+\lambda I,
$$
for some non-self-adjoint operator $I$ related to $V$, see \fer{4.1}. We present the explicit construction of $K_\lambda$ in Appendices \ref{appgluing} and \ref{appA}. It gives that $\e^{\i tL_\lambda} A\e^{-\i tL_\lambda} = \e^{\i tK_\lambda} A\e^{-\i tK_\lambda}$, and that $\e^{-\i tK_\lambda}\Omega_{\beta,0}=\Omega_{\beta,0}$. 
 Using these relations we obtain from \fer{3}
\begin{equation}
\av{A}_t =\scalprod{(B^*\psi_0)\otimes\Omega_{\r,\beta}}{\e^{\i tK_\lambda} A \Omega_{\beta,0}}.
\label{100}
\end{equation}
The operator $K_\lambda$ is unbounded and non-self-adjoint, and we do not know a priori whether $\e^{\i tK_\lambda}$ is defined on $A\Omega_{\beta,0}$. Consequently, the formula above and the next one are formal, heuristic expressions (which obtain a rigorous meaning using a complex deformation, see below and \cite{MMS1,MMS2}). 
Furthermore, we represent the propagator as an integral over the resolvent, \cite{MMS1,MMS2} to arrive at
\begin{equation}
\av{A}_t = \frac{-1}{2\pi\i} \int_{\rx-\i}  \e^{\i tz} \scalprod{(B^*\psi_0)\otimes\Omega_{\r,\beta}}{(K_\lambda-z)^{-1} A \Omega_{\beta,0}} \d z.
\label{6}
\end{equation}

Next we ``uncover the resonances''. We use the following notation for the second quantization of a a one-body operator $O$ acting on single-particle wave functions of the variable $k\in\rx^3$: 
\begin{equation}
\d\Gamma(O) = \int_{\rx^3}  a^*(k) O a(k)\, \d^3k.
\label{xxx}
\end{equation}
Let 
$$
N=\d\Gamma(\bbbone)\otimes\bbbone_\r + \bbbone_\r\otimes\d \Gamma(\bbbone)
$$
be the {\it total} number operator on $\h_\r={\cal F}(L^2(\rx^3,\d^3k))\otimes{\cal F}(L^2(\rx^3,\d^3k))$, and define the operator
\begin{equation}
D=\d\Gamma(\vartheta)\otimes\bbbone - \bbbone_\r\otimes \d\Gamma(\vartheta),
\label{opD}
\end{equation}
where $\vartheta= \frac \i 2(\hat k\cdot\nabla + \nabla\cdot \hat k)$, with $\hat k=\frac{k}{|k|}$. The operator $D$ is self-adjoint (even though $\d\Gamma(\vartheta)$ alone is only symmetric but not self-adjoint, see \cite{MMS1}). Let 
\begin{equation}
U(\omega)=\e^{-\i\omega D}, 
\label{Uomega}
\end{equation}
$\omega\in\rx$, be the group of unitary transformations on $\h_\r$ generated by $D$. We transform the operator $K_\lambda$ unitarily as $K_\lambda(\omega)=U(\omega)K_\lambda U(\omega)^{-1}$. From $-\i[\d\Gamma(\vartheta),\d\Gamma(|k|)]=\d\Gamma(\bbbone)$, and from equations \fer{69}, \fer{70} and \fer{75}, we obtain the relation $\i[D, L_0]=N$. Consequently,
\begin{equation}
K_\lambda(\omega) = L_0+\omega N +\lambda I(\omega),
\label{15}
\end{equation}
where $I(\omega)=U(\omega) I U(\omega)^{-1}$. The explicit formula is presented in Appendices \ref{appgluing} and \ref{appA}. Since $U(\omega)$ is unitary (for $\omega\in\rx$), and since both $A\Omega_{\beta,0}$ and $(B^*\psi_0)\otimes\Omega_{\r,\beta}$ are invariant under $U(\omega)$, the r.h.s. of equation \fer{6} equals that of \fer{16}, for all $\omega\in\rx$. The integrand in \fer{16} can be extended to an analytic function in $\omega$, for $\omega$ in a strip $0<{\rm Im} \omega<\omega_0=\frac{2\pi}{\beta}$ (which is continuous as ${\rm Im}\omega\downarrow 0$). Since the r.h.s. of \fer{16} is constant in $\omega$ for real $\omega$, it follows that \fer{16} holds {\it for all complex $\omega$ in the strip}.  This concludes our heuristic derivation of formula \fer{16}.

\medskip
Let us take $\omega=\i\omega'$, for some $\omega'>0$. The point is that the spectrum of $K_\lambda(\omega)$ is much easier to analyze than that of $K_\lambda$.  This is so because the spectrum of $K_0(\omega)=L_0+\i \omega' N$ is
\begin{equation}
{\rm spec}(K_0(\omega)) = \left(\{E_i-E_j\}_{i,j=1,\ldots,N}\right)\cup_{n\geq 1}  ( \i \omega'  n +\rx).
\label{sepgap}
\end{equation}
The eigenvalues $E_i-E_j$ have eigenvectors $\varphi_i\otimes\varphi_j\otimes\Omega_{\r,\beta}$ and the lines $\i\omega' n+{\mathbb R}$ are horizontal branches of continuous spectrum. There is a gap of size $\omega'$ separating the eigenvalues from the continuous spectrum of $K_0(\omega)$, so if we add the perturbation $\lambda I(\omega)$, which is bounded relative to $K_0(\omega)$, we can follow the location of eigenvalues by simple (analytic) perturbation theory, provided $\lambda$ is small compared to $\omega'$. One obtains the following result.

\begin{theorem}
\label{lemma}
Fix $\omega'>0$. There is a constant $c_0>0$ s.t. if $|\lambda|\leq c_0/\beta$ then, for all $\omega$ with ${\rm Im}\omega>\omega'$, the spectrum of $K_\lambda(\omega)$ in the complex half-plane $\{{\rm Im}z < \omega'/2\}$ is independent of $\omega$ and consists purely of the distinct eigenvalues
$$
\{\varepsilon_e^{(s)}(\lambda)\ |\ e\in{\rm spec}(L_\s), s=1,\ldots,\nu(e)\},
$$
where $1\leq \nu(e)\leq {\rm mult}(e)$ counts the splitting of the eigenvalue $e$. Moreover, we have $\lim_{\lambda\rightarrow 0}|\varepsilon_e^{(s)}(\lambda)-e|=0$ for all $s=1,\ldots,\nu(e)$, and we have ${\rm Im}\varepsilon_e^{(s)}(\lambda)\geq 0$. Also, the continuous spectrum of $K_\lambda(\omega)$ lies in the region $\{{\rm Im}z \geq 3\omega'/4\}$.
\end{theorem}

{\it Remarks.\ } 1.\ The proof of the theorem uses standard analytic perturbation theory. The constant $c_0$ in Theorem \ref{lemma} is defined by the condition that the eigenvalues $\varepsilon_e^{(s)}(\lambda)$ stay away from the continuous spectrum of $K_\lambda(\omega)$. Since 

(a) the continuous spectrum of $K_\lambda(\omega)$ lies in a neigbourhood of order $\lambda$ around the continuous spectrum of $K_0(\omega)$, i.e., in the region $\{{\rm Im}z\geq 3\omega'/4\}$, provided $|\lambda|\leq c_1\omega'$, for some constant $c_1$ (see \fer{sepgap}), and 

(b) the eigenvalues $\varepsilon_e^{(s)}(\lambda)$ have imaginary part bounded from above by $c_2\lambda^2$, for some constant $c_2$, see \fer{int1},

\noindent
the eigenvalues and the continuous spectrum of $K_\lambda(\omega)$ are separated provided $|\lambda|\leq \min\{c_1,c_2\}\omega'<2\pi \min\{c_1,c_2\}/\beta$.

The fact that the spectrum of $K_\lambda(\omega)$ must lie in the closed upper half plane is quite clear since $\av{A}_t$, \fer{6}, must stay bounded as $t\rightarrow\infty$.

2.\ By construction we have $K_\lambda(\omega)\Omega_{\beta,0}=0$, so we set $\varepsilon_0^{(1)}=0$.

\medskip

We now perform the {\it pole approximation} by deforming the contour $z=\rx-\i$ in the integral \fer{16} into the contour $z=\rx+\frac{\i}{2} [\mu +\omega'/2]$, where we introduce
\begin{equation}
\mu= \max_{e,s}\{{\rm Im}\, \varepsilon_e^{(s)}(\lambda)\}.
\label{maxim}
\end{equation}
This contour separates the eigenvalues $\varepsilon_e^{(s)}(\lambda)$ (lying below it) from the continuous spectrum (lying above it). Indeed, the gap between the eigenvalue with biggest imaginary part and the continuous spectrum is bounded from below by $3\omega'/4-\mu$.

With this contour deformation, we pick up the residues of the poles of the integrand, sitting at the resonance energies $\varepsilon_e^{(s)}(\lambda)$. Let $\cc_e^{(s)}$ be a small circle around $\varepsilon_e^{(s)}$ not enclosing any other point of the spectrum of $K_\lambda(\omega)$. Then \fer{16} gives
\begin{eqnarray}
\lefteqn{\av{A}_t= \label{17}}\\
&& \frac{-1}{2\pi\i} \sum_e\sum_{s=1}^{\nu(e)} \int_{\cc_e^{(s)}}  \e^{\i tz} \scalprod{(B^*\psi_0)\otimes\Omega_{\r,\beta}}{(K_\lambda(\omega)-z)^{-1} A \Omega_{\beta,0}} \d z +R,\nonumber
\end{eqnarray}
where
\begin{equation}
R=\frac{-1}{2\pi\i}\int_{\rx+\frac{\i}{2}[\mu+\omega'/2]}  \e^{\i tz} \scalprod{(B^*\psi_0)\otimes\Omega_{\r,\beta}}{(K_\lambda(\omega)-z)^{-1} A \Omega_{\beta,0}} \d z.\ \ \ \
\label{17.1}
\end{equation}
We prove the following estimate on the remainder term $R$ in Appendix \ref{prooflemmalso}.

\begin{proposition}
\label{propremainder}
We have $R = O(\lambda^2\e^{-\frac t2 [\mu+ \omega'/2]})$.
\end{proposition}

For simplicity of the exposition, we assume in the remainder of this section that the nonzero resonance energies $\varepsilon_e^{(s)}\neq 0$ are simple poles of the resolvent $(K_\lambda(\omega)-z)^{-1}$ (i.e., that the $\varepsilon_e^{(s)}$ are semisimple eigenvalues of $K_\lambda(\omega)$; this will be satisfied in all our applications). It is easy to extend the following arguments to the general situation.

Under the assumption above we can replace $\e^{\i tz}$ by $\e^{\i t\varepsilon_e^{(s)}}$ in the first term on the r.h.s. of \fer{17},
\begin{equation}
\av{A}_t=
\sum_e\sum_{s=1}^{\nu(e)} \e^{\i t\varepsilon_e^{(s)}} \scalprod{(B^*\psi_0)\otimes\Omega_{\r,\beta}}{Q_e^{(s)} A \Omega_{\beta,0}} + O(\lambda^2\e^{-\omega' t/2}),
\label{18}
\end{equation}
where we introduced the (in general, non-orthogonal) projections
$$
Q_e^{(s)}=Q_e^{(s)}(\omega) =\frac{-1}{2\pi\i}\int_{\cc_e^{(s)}} (K_\lambda(\omega)-z)^{-1}\d z.
$$
If $\varepsilon_e^{(s)}$ is a simple eigenvalue of $K_\lambda(\omega)$, then we have
$$
Q_e^{(s)} = |\chi_e^{(s)}\rangle\langle\widetilde\chi_e^{(s)}|,
$$
where the vectors $\chi_e^{(s)}$ and $\widetilde\chi_e^{(s)}$ satisfy
\begin{equation}
K_\lambda(\omega)\chi_e^{(s)} = \varepsilon_e^{(s)}\chi_e^{(s)} \mbox{\ \ and\ \ } (K_\lambda(\omega))^*\widetilde\chi_e^{(s)} = \overline{\varepsilon_e^{(s)}}\widetilde\chi_e^{(s)},
\label{4.21a}
\end{equation}
and are normalized as
\begin{equation}
\scalprod{\chi_e^{(s)}}{\widetilde\chi_e^{(s)}}=1.
\label{40}
\end{equation}
We obtain from \fer{18} the relation
\begin{equation}
\av{\av{A}}_\infty:=\lim_{T\rightarrow\infty} \frac 1T\int_0^T \av{A}_t\d t = \sum_{s':\ \varepsilon_0^{(s')}= 0} \scalprod{(B^*\psi_0)\otimes\Omega_{\r,\beta}}{Q_0^{(s')} A\Omega_{\beta,0}}.
\label{19}
\end{equation}
All the other terms vanish in the ergodic mean limit.

If $\av{A}_t$ has a limit as $t\rightarrow\infty$, as is the case when ${\mathrm Im}\,\varepsilon_e^{(s)}(\lambda)>0$ for $\varepsilon_e^{(s)}(\lambda)\neq 0$, then $\av{\av{A}}_\infty$ is just that limit (it may happen that $\av{A}_t$ does not have a limit, but $\av{\av{A}}_\infty$ always exists as we see from \fer{19}). We have thus identified the limit term in the expansion \fer{18} and we obtain the results of Theorem \ref{prop3.2}.

For specific models, one can calculate (perturbatively in $\lambda$, to any order) the resonance energies $\varepsilon_e^{(s)}$ and the projection operators $Q_e^{(s)}$, and one obtains estimates on the difference $\av{A}_t - \av{\av{A}}_\infty$.

{\it Remark.\ } As the results above show, the long time behaviour of averages $\av{A}_t$ is determined by the resonance energies $\varepsilon_e^{(s)}$ and the resonance eigenvectors $\chi_e^{(s)}$ and $\widetilde\chi_e^{(s)}$ of the family $K(\omega)$ and its adjoint. This resonance data can be constructed from eigenvalues and eigenfunctions of operators $\Lambda_{e,\beta,\lambda}$ which act on $\hh_\s\otimes\hh_\s$ and which are independent of $\omega$. (This reconstruction can be viewed as a kind of ``inversion'' of the Feshbach map, see Sections \ref{sect} and \ref{rrrsect}.) We point out that this procedure is relatively easy to implement for translation-analytic systems, i.e., systems satisfying Condition A stated after \fer{7}. For systems wich do not possess this condition a renormalization group analysis in the spirit of \cite{BFS1,BFS2} has to be carried out, see also \cite{MMS1,MMS2}.

\section{Resonance energies $\varepsilon_e^{(s)}(\lambda)$}
\label{sect}

The goal of this section is to evaluate the main contribution (in $\lambda$) to the resonance energies $\varepsilon_e^{(s)}(\lambda)$, where $e\in\{E_m-E_n\ |\ m,n=1,\ldots,N\}$. Using a standard Feshbach-map argument (see e.g. \cite{BFS,BCFS,MMS1,MMS2}) we obtain the expansion
\begin{equation}
\varepsilon_e^{(s)}(\lambda) = e-\lambda^2 \delta_e^{(s)} + O(\lambda^4),
\label{26}
\end{equation}
where the $\delta_e^{(s)}\in\cx$ are the eigenvalues of the so-called {\it level shift operator} $\Lambda_e$ associated to $e$, defined by
\begin{equation}
\Lambda_e = P_e I \overline P_e(\overline L_0 -e +\i 0)^{-1} \overline P_e I P_e,
\label{27}
\end{equation}
where $P_e=P(L_\s=e)\otimes P_\Omega$ is the eigen-projection of $L_0$ associated to the eigenvalue $e$, $\overline P_e=\bbbone-P_e$, and the restriction of $\overline P_e L_0\overline P_e$ to the range of $\overline P_e$ is denoted by $\overline L_0$ (see also \cite{Mlso}).

Consider an interaction of the form \fer{31}, where $G$ is a symmetric matrix on $\hh_\s=\cx^n$.  The following result follows from a direct calculation involving the explicit form of $\Lambda_e$. We prove it in Appendix \ref{prooflemmalso}.

\begin{proposition}[Level shift operators]
\label{LSO}
The operator
$$
\Lambda_e(\epsilon):= P_e I \overline P_e(\overline L_0 -e +\i \epsilon)^{-1} \overline P_e I P_e
$$
has the representation
\begin{equation}
\Lambda_e(\epsilon)= \Lambda_{e,{\mathrm d}}(\epsilon) + \Lambda_{e,{\mathrm o}}(\epsilon) +\Lambda_{e,{\mathrm m}}(\epsilon),
\label{lso1}
\end{equation}
where the subscripts ``{\rm d,o,m}'' stand for ``{\rm diagonal, off-diagonal, mixed}'' and refer to the decomposition $G=G_{\mathrm d}+G_{\mathrm o}$ into a sum of a diagonal and an off-diagonal matrix in the energy basis. Let $\overline G$ be the matrix obtained by taking the complex conjugate of $G$, in the energy basis. We have
\begin{eqnarray}
\lefteqn{
2\Lambda_{e,{\mathrm d}}(\epsilon)}\label{20}\\
 &=& P_e (G_{\mathrm d}\otimes\bbbone -\bbbone\otimes \overline G_{\mathrm d})
(G_{\mathrm d}\otimes\bbbone +\bbbone\otimes \overline G_{\mathrm d})
P_e \ \scalprod{g}{\frac{\omega}{\omega^2+\epsilon^2}g}\nonumber\\
&&- P_e \left(G_{\mathrm d} \otimes\bbbone -\bbbone\otimes \overline G_{\mathrm d}\right)^2P_e \ \scalprod{g}{\coth\!\!\left(\frac{\beta\omega}{2}\right)\frac{\i\epsilon}{\omega^2+\epsilon^2}g},\ \
\nonumber
\end{eqnarray}
\begin{eqnarray}
\lefteqn{2\Lambda_{e,{\mathrm o}}(\epsilon)}
\label{21}\\
 &=& P_e (G_{\mathrm o} \otimes\bbbone) \int_{\rx\times S^2} \frac{u^2|g(|u|,\sigma)|^2}{|1-\e^{-\beta u}|}(L_\s-e+u+\i\epsilon)^{-1}\ (G_{\mathrm o}\otimes\bbbone)P_e\nonumber\\
&+& P_e (\bbbone\otimes \e^{-\frac\beta 2H_\s}\overline G_{\mathrm o}) \int_{\rx\times S^2}\frac{u^2|g(|u|,\sigma)|^2}{|1-\e^{+\beta u}|}(L_\s-e+u+\i\epsilon)^{-1}\ (\bbbone\otimes \overline G_{\mathrm o}\e^{\frac\beta 2H_\s})P_e\nonumber\\&-& P_e (G_{\mathrm o}\otimes \bbbone) \int_{\rx\times S^2}\frac{u^2|g(|u|,\sigma)|^2}{|1-\e^{-\beta u}|}(L_\s-e+u+\i\epsilon)^{-1}\ (\bbbone\otimes \e^{-\frac\beta 2H_\s}\overline G_{\mathrm o}\e^{\frac\beta 2H_\s})P_e\nonumber\\
&-& P_e (\bbbone\otimes \e^{-\frac\beta 2 H_s}\overline G_{\mathrm o}\e^{\frac\beta 2H_\s}) \int_{\rx\times S^2}\frac{u^2|g(|u|,\sigma)|^2}{|1-\e^{+\beta u}|}(L_\s-e+u+\i\epsilon)^{-1}\ (G_{\mathrm o}\otimes \bbbone)P_e\nonumber
\end{eqnarray}
\begin{eqnarray}
\lefteqn{
2\Lambda_{e,{\mathrm m}}(\epsilon)}\label{22}\\
 &=& P_e \Big\{ (G_{\mathrm d}\otimes\bbbone +\bbbone\otimes \overline G_{\mathrm d})\ ,\
(G_{\mathrm o}\otimes\bbbone -\bbbone\otimes \e^{-\frac\beta 2H_\s}\overline G_{\mathrm o}\e^{\frac\beta 2 H_\s})\Big\}P_e \times\nonumber\\
& &\times \scalprod{g}{\frac{\omega}{\omega^2+\epsilon^2}g}\nonumber\\
&-& P_e \Big\{ (G_{\mathrm d} \otimes\bbbone -\bbbone\otimes \overline G_{\mathrm d}) \ ,\ (G_{\mathrm o}\otimes\bbbone - \bbbone\otimes \e^{-\frac\beta 2H_\s}\overline G_{\mathrm o}\e^{\frac\beta 2 H_\s}) \Big\} P_e \times\nonumber\\
&&\times \scalprod{g}{\coth\!\!\left(\frac{\beta\omega}{2}\right)\frac{\i\epsilon}{\omega^2+\epsilon^2}g},\nonumber
\end{eqnarray}
where $\{A,B\}=AB+BA$ is the anti-commutator.
\end{proposition}

{\it Remarks.\ }
1. If $e$ is a simple eigenvalue then all factors $\e^{\pm\frac\beta 2H_\s}$ in \fer{20}-\fer{22} can be set equal to one.

2. If the eigenvalues of $H_\s$ are non-degenerate then $\Lambda_0(\epsilon) = \Lambda_{0,{\mathrm o}}(\epsilon)$. In particular, the eigenvalues $\delta_0^{(s)}$ do not depend on the diagonal elements of $G$.

To see that the statements in the remark are true, simply use in formulas \fer{20}-\fer{22}, the facts that $P_0 = \sum_j p_j\otimes p_j\otimes P_\r$, where $p_j=|\varphi_j\rangle\langle\varphi_j|$ and $P_\r=|\Omega_{\r,\beta}\rangle\langle\Omega_{\r,\beta}|$, and that $P_e=p_i\otimes p_j\otimes P_\r$ if $e$ is simple.

\section{Resonance eigenvectors  $\chi_e^{(s)}(\lambda)$ and $\widetilde\chi_e^{(s)}(\lambda)$}
\label{rrrsect}

In this section we derive the following expansions of $\chi_e^{(s)}(\lambda)$ and $\widetilde\chi_e^{(s)}(\lambda)$ (defined in \fer{4.21a}) in $\lambda$:
\begin{eqnarray}
\chi_e^{(s)} &=& \left[\bbbone - \lambda \overline P_e (\overline L_0(\omega)-e)^{-1}\overline P_e I(\omega) P_e \right] \eta_e^{(s)}\otimes\Omega_{\r,\beta} +O(\lambda^2), \label{30}\\
\widetilde\chi_e^{(s)} &=& \left[\bbbone - \lambda \overline P_e(\overline L_0(\overline\omega)-e)^{-1}\overline P_e (I^*)(\overline\omega) P_e \right] \widetilde\eta_e^{(s)}\otimes\Omega_{\r,\beta} +O(\lambda^2),
\label{30.1}
\end{eqnarray}
where $\eta_e^{(s)}\otimes\Omega_{\r,\beta}$ and $\widetilde\eta_e^{(s)}\otimes\Omega_{\r,\beta}$ denote the eigenvectors of the level shift operator $\Lambda_e$ and its adjoint $(\Lambda_e)^*$, respectively,
\begin{eqnarray}
\Lambda_e\ \eta_e^{(s)}\otimes\Omega_{\r,\beta} &=& \delta_e^{(s)}\  \eta_e^{(s)}\otimes\Omega_{\r,\beta},\label{39}\\
(\Lambda_e)^*\ \widetilde\eta_e^{(s)}\otimes\Omega_{\r,\beta} &=& \overline{\delta_e^{(s)}}\  \widetilde\eta_e^{(s)}\otimes\Omega_{\r,\beta}.
\label{39.1}
\end{eqnarray}
To arrive at the expansions \fer{30}, \fer{30.1}, we use the method of the Feshbach map (see e.g. \cite{BFS1,BFS2}), according to which we know that
\begin{equation}
\chi_e^{(s)} = \left( P_e - \lambda \overline P_e (\overline K_\lambda(\omega)-\varepsilon_e^{(s)})^{-1} \overline P_e I(\omega)\right) P_e \xi_e^{(s)},
\label{28}
\end{equation}
where $\overline K_\lambda(\omega) =\overline P_e K_\lambda(\omega) \overline P_e\upharpoonright {\rm Ran}\overline P_e$, and 
where $\xi_e^{(s)}\in {\mathrm Ran}P_e$ is the eigenvector of
\begin{equation}
F_{P_e,\varepsilon_e^{(s)} }(K_\lambda(\omega)):= P_e\left( e -\lambda^2 I(\omega)\overline P_e(\overline K_\lambda(\omega) -\varepsilon_e^{(s)})^{-1}\overline P_e I(\omega)\right) P_e,
\label{29}
\end{equation}
with eigenvalue $\varepsilon_e^{(s)}$. We expand the r.h.s. of \fer{29} in $\lambda$ using a Neumann series expansion of the resolvent $\overline P_e(\overline K_\lambda(\omega) -\varepsilon_e^{(s)})^{-1}\overline P_e= \overline P_e(\overline K_0(\omega) +\lambda\overline I(\omega) -\varepsilon_e^{(s)}(\lambda))^{-1}\overline P_e$. This gives
$$
F_{P_e,\varepsilon_e^{(s)}}(K_\lambda(\omega))= P_e \big(e -\lambda^2\Lambda_e + O(\lambda^4)\big)P_e,
$$
where we also used \fer{26}. Consequently, taking into account the definition \fer{39}, we get the expansion
$$
\xi_e^{(s)} = \eta_e^{(s)}\otimes\Omega_{\r,\beta} +O(\lambda^2).
$$
Finally, a similar expansion of the r.h.s. of \fer{28} gives \fer{30}.

The analogous expansion \fer{30.1} for $\widetilde\chi_e^{(s)}$ is obtained by proceding as above, and by using the following little result.
\begin{lemma}
\label{littlelemma}
The level shift operator associated to the eigenvalue $e$ and the adjoint operator $(K_\lambda(\omega))^*$, ${\mathrm Im}\omega>0$,  is given by
$$
P_e (I^*)(\overline\omega) \overline P_e (\overline L_0(\overline \omega)-e)^{-1}\overline P_e (I^*)(\omega) P_e = (\Lambda_e)^*.
$$
\end{lemma}

{\it Proof.\ }
We have (recall \fer{opD} and \fer{Uomega})
$$
(K_\lambda(\omega))^* = [\e^{-i\omega D} K_\lambda\e^{\i\omega D}]^*
=\e^{-\i\overline \omega D} K_\lambda^* \e^{\i \overline\omega D}
=
L_0(\overline\omega) +\lambda (I^*)(\overline\omega).
$$
Therefore we have the following expression for the level shift operator,
\begin{eqnarray*}
\lefteqn{
P_e (I^*)(\overline\omega) \overline P_e (\overline L_0(\overline \omega)-e)^{-1}\overline P_e (I^*)(\overline\omega) P_e}\\
&& = P_e I^* \overline P_e (\overline L_0-e-\i 0)^{-1}\overline P_e I^* P_e\\
&& = \left[ P_e I \overline P_e (\overline L_0-e+\i 0)^{-1}\overline P_e I P_e\right]^*\\
&& = (\Lambda_e)^*.
\end{eqnarray*}
In order to get rid of the parameter $\overline\omega$ by analyticity, we have introduced in the first step the term $-\i0$. \hfill $\blacksquare$

\section{Special cases}
\label{sect4}

\subsection{The qubit: proof of Theorem \ref{thm1}}
\label{subsqubit}
In this section we calculate the level shift operators and their spectral data for a qubit coupled to the Bose field. Recall that a qubit is a two-dimensional system, with state space (of pure states) $\hh_\s=\cx^2$, and Hamiltonian $H_\s={\rm diag}(E_1,E_2)$. The interaction to the Bose field is given by \fer{michael*} (which becomes \fer{31} in the Hilbert space of positive temperature states). Recall the definition of the coupling parameter $\xi(\eta)$, \fer{35}. We set
$$
\varphi_1=
\left[
\begin{array}{c}
1\\
0
\end{array}
\right],\ \ \
\varphi_2=
\left[
\begin{array}{c}
0\\
1
\end{array}
\right].
$$
Let $\Delta=E_2-E_1>0$ denote the gap in the energy spectrum of $H_\s$. The following result is an easy application of Proposition \ref{LSO} to the specific model.
\begin{proposition}
\label{qubitlso}
In the basis $\{\varphi_1\otimes\varphi_1,\varphi_2\otimes\varphi_2\}$, the level shift operator $ \Lambda_0$ is given by
\begin{equation}
2\Lambda_0 = -\i\pi^2|c|^2\frac{\xi(\Delta)}{\cosh(\beta\Delta/2)}
\left[
\begin{array}{cc}
\e^{-\beta\Delta/2} & -1 \\
-1 & \e^{\beta\Delta/2}
\end{array}
\right].
\label{36}
\end{equation}
The (one-dimensional) level shift operator $\Lambda_\Delta$ is given by
\begin{eqnarray}
2\Lambda_\Delta &=& (b^2-a^2) \scalprod{g}{\omega^{-1}g} +|c|^2 {\rm P.V.}\int_{\rx\times S^2} u^2|g(|u|,\sigma)|^2 \coth\!\left(\frac{\beta |u|}{2}\right)\frac{1}{u-\Delta}\nonumber\\
&&-\i\pi^2|c|^2\xi(\Delta) -\i\pi (b-a)^2\xi(0),
\label{37}
\end{eqnarray}
and $\Lambda_{-\Delta}$ is obtained from the r.h.s. of \fer{37} by switching the sign of the real part, $\Lambda_{-\Delta}=-\overline{\Lambda_{\Delta}}$.
\end{proposition}
{\it Remarks.\ }
1.\ The vector $\Omega_{\s,\beta}\propto \varphi_1\otimes\varphi_1 + e^{-\beta\Delta/2}\varphi_2\otimes\varphi_2$ spans the kernel of $\Lambda_0$. The gap in the spectrum of $\Lambda_0$ is $\pi^2|c|^2\xi(\Delta)$, it is exactly twice the gap of ${\mathrm Im}\Lambda_{\pm\Delta}$ coming from a $G$ with constant diagonal in the energy basis.

2.\  $\xi(0)$ is non-zero only if the infra-red behaviour of $g$ is given by $p=-1/2$, c.f. \fer{38}.

\medskip

It is also easy to obtain the eigenvectors of the level shift operators (and their adjoints, see Lemma \ref{littlelemma}).

\begin{proposition}
\label{lemma3}
For the qubit model, the vectors determined by \fer{39} and \fer{39.1} are
\begin{eqnarray*}
\eta_0^{(1)}&=&\widetilde\eta_0^{(1)} = \Omega_{\s,\beta} = (1+\e^{-\beta\Delta})^{-1/2} \left(\varphi_1\otimes\varphi_1 + \e^{-\beta\Delta/2}\varphi_2\otimes\varphi_2\right),\\
\eta_0^{(2)}&=&\widetilde\eta_0^{(2)} = (1+\e^{-\beta\Delta})^{-1/2}\left(\e^{-\beta\Delta/2}\varphi_1\otimes\varphi_1 -  \varphi_2\otimes\varphi_2\right),\\
\eta_\Delta &=& \widetilde\eta_\Delta = \varphi_2\otimes\varphi_1,\\
\eta_{-\Delta} &=&  \widetilde\eta_{-\Delta} =\varphi_1\otimes\varphi_2,
\end{eqnarray*}
they are normalized as $\scalprod{\eta_e^{(s)}}{\widetilde\eta_e^{(s)}}=1$, as required by \fer{40}.
\end{proposition}

Propositions \ref{qubitlso} and \ref{lemma3} together with Theorem \ref{prop3.2} imply Theorem \ref{thm1}. \hfill $\blacksquare$

\medskip
\noindent
Let us show the assertion of Remark 5 after \fer{i4}. Suppose that $\rhobar_0=|\varphi_j\rangle\langle\varphi_j|$ for a fixed $j=1,2$. This initial state is represented on $\h_\s=\cx^2\otimes\cx^2$ by the vector $\psi_0=\varphi_j\otimes\varphi_j$. The operator $B$ defined by relation \fer{96} is $B=c_j\bbbone_\s\otimes |\varphi_j\rangle\langle\varphi_j|$, where $c_j$ is a complex number. Thus $B^*\psi_0=\overline c_j\psi_0$ and we may replace $B^*$ by $\overline c_j$ in formula \fer{24}, Theorem \ref{prop3.2}. By Proposition \ref{lemma3} we have $Q_{\pm\Delta} = |\eta_{\pm\Delta}\rangle\langle\widetilde\eta_{\pm\Delta}|\otimes|\Omega_{\r,\beta}\rangle\langle\Omega_{\r,\beta}|+O(\lambda^2)$, so the terms in \fer{24} coming from the nonzero resonances are
$$
c_j\scalprod{\psi_0\otimes\Omega_{\r,\beta}}{Q_{\pm\Delta} A\Omega_{\beta,0}}
=c_j\scalprod{\varphi_j\otimes\varphi_j\otimes\Omega_{\r,\beta}}{Q_{\pm\Delta} A \Omega_{\beta,0}} = O(\lambda^2),
$$
since $\scalprod{\varphi_j\otimes\varphi_j}{\eta_{\pm\Delta}}=0$. The calculation of the constants $C_0$ for $j=1,2$ is carried out in the same fashion. Here one uses that $A=p_{1,1}\otimes\bbbone_\s\otimes\bbbone_\r$ and the explicit form of $\Omega_{\beta,0}$, see \fer{3.1} and \fer{62}.

Finally we show how to arrive at the relations given in the illustration involving \fer{illust}. As one easily verifies the state $\psi_0$ representing \fer{illust} in $\hh_\s\otimes\hh_\s$ is given by $\psi_0=\frac{1}{\sqrt{2}}(\varphi_1\otimes\varphi_2+\varphi_2\otimes\varphi_2)$. The associated operator $B$ (see \fer{96}) is
$$
B=\sqrt{\frac{Z_{\s,\beta}}{2}}\  \bbbone_\s\otimes
\left[
\begin{array}{cc}
0 & 0 \\
\e^{\beta E_1/2} & \e^{\beta E_2/2}
\end{array}
\right].
$$
The relations after \fer{illust} follow by easy direct calculation, using Theorem \ref{prop3.2} and Propositions \ref{qubitlso} and \ref{lemma3}.

\subsection{Energy conserving interactions}
\label{sectapp2}

In this section we apply the dynamical resonance theory to interactions which commute with $H_\s$. Such quantum non-demolition interactions are widely studied in the literature, see e.g. \cite{DBV,DG,PSE,MP,VK} and references therein (this is a small fraction of works on this subject -- there is an immense number of papers on it). Some energy conserving models can be solved explicitly and provide benchmark examples for our methods, as well as starting points for a perturbative treatment (for small $[H_\s,v]$).

\bigskip
Let $\hh_\s=\cx^N$, $H_\s={\rm diag}(E_1,\ldots,E_N)$, where $E_1 < E_2 < \cdots < E_N$. (A similar analysis can be carried out if there are degenerate eigenvalues.)  We couple $\s$ to the bosonic reservoir via the interaction $\lambda v$, \fer{michael*}, which is represented on the positive temperature Hilbert space by the operator $\lambda V$ of the form \fer{31}, with a matrix $G$ that commutes with $H_\s$, i.e.,
$$
G={\rm diag}(\gamma_1,\ldots,\gamma_N),
$$
with $\gamma_j\in\rx $. Let $d\geq 1$ be the dimension of the quantum field, and $\omega=|k|$. The dynamical resonance theory yields the following result (which we prove in Appendix \ref{prooflemmalso}).

\begin{proposition}
\label{comparison}
Theorem \ref{prop3.2} implies that $[\rhobar_t]_{m,m}=[\rhobar_{0}]_{m,m}$, and that, for $m\neq n$,
\begin{equation}
[\rhobar_t]_{m,n} = ([\rhobar_0]_{m,n}+O(\lambda^2)) \e^{-\i t(E_m-E_n) +\i\lambda^2 t [\delta_{E_n-E_m}+O(\lambda^2)]} + O(\lambda^2\e^{-t\omega'/2}),
\label{q4}
\end{equation}
where
\begin{equation}
\delta_{E_n-E_m} =\textstyle{\frac 12}(\gamma_m^2-\gamma_n^2)\scalprod{g}{\omega^{-1}g}+\i(\gamma_m-\gamma_n)^2
\left\{
\begin{array}{ll}
0 & \mbox{if $p>\frac{2-d}{2}$}\\
\pi\xi(0)>0 &\mbox{if $p=\frac{2-d}{2}$}\\
+\infty & \mbox{if $p<\frac{2-d}{2}$}
\end{array}
\right.
\label{lim1}
\end{equation}
Here, $p$ is the power characterizing the infrared behaviour of $g$, c.f. \fer{38}, and $\xi(0)$ is given by \fer{35} in which the integral is taken over $\rx^d$ instead of $\rx^3$.
\end{proposition}

As it turns out, this model is explicitly solvable.
For illustration purposes we compare the explicit solution to the results obtained by our dynamical resonance method.

\begin{proposition}[Explicit solution]
\label{prop7.1}
The reduced density matrix elements are given by
\begin{equation*}
[\rhobar_t]_{m,n} =  [\rhobar_0]_{m,n}\e^{-\i t(E_m-E_n)+\i\lambda^2\alpha_{m,n}(t)},
\end{equation*}
with $\alpha_{m,n}(t) =  (\gamma_m^2-\gamma_n^2) S(t)+\i (\gamma_m-\gamma_n)^2 \Gamma(t)$, where
\begin{eqnarray*}
\Gamma(t) &=&  \int_{\rx^d} |g(k)|^2\coth\left(\frac{\beta\omega}{2}\right)\frac{\sin^2(\frac{\omega t}{2})}{\omega^2}\d^dk,\\
S(t) &=& \frac 12\int_{\rx^d} |g(k)|^2 \ \frac{\omega t-\sin\omega t}{\omega^2}\d^dk,
\end{eqnarray*}
and where $\omega(k)=|k|$, $\sigma(k)\in S^{d-1}$. 
\end{proposition}

{\it Remarks.\ }  1)\ The same result for the decoherence function $\Gamma(t)$ has been obtained in \cite{PSE,MP}. There, the strategy was to obtain an explicit expression for the matrix elements, first for the system where the reservoir's modes are discrete, and to take afterwards the continuous mode limit. Our calculation is performed directly on the system with continuous modes. We give it in Appendix \ref{prooflemmalso}. 

2)\ We have $\scalprod{g}{\omega^{-1}g}<\infty$, as follows from the assumption $\mu_\beta g\in L^2(\rx^d,\d^dk)$. The latter is required for the interaction \fer{michael*}, \fer{83} to be defined (recall that $\mu_\beta$ has an $\omega^{-1/2}$-singularity at zero, see \fer{planck}).

 The next statement is evident from the explicit form of the matrix elements.
\begin{corollary}
\label{corollary11}
The populations are constant, $[\rhobar_t]_{m,m}= [\rhobar_0]_{m,m}$ for all $m$ and all $t$. Full decoherence takes place if and only if $\Gamma(t)\rightarrow\infty$ as $t\rightarrow\infty$, i.e., if and only if $p\leq \frac{2-d}{2}$. 
\end{corollary}

The following result examines the asymptotic behaviour of $\alpha_{m,n}(t)$. It illustrates the compatibility between the explicit solution and the result obtained from the resonance theory. Its proof follows readily from the explicit expressions given in Proposition \ref{prop7.1}.
\begin{proposition}
\label{ppropp}
Let $\delta_{E_n-E_m}$ be defined as in \fer{lim1}. We have
$$
\lim_{t\rightarrow\infty}\frac{\alpha_{m,n}(t)}{t} = \delta_{E_n-E_m}.
$$
For $p=\frac{2-d}{2}$ the decoherence function grows asymptotically linearly in time, $\Gamma(t)\sim\pi\xi(0)t$, for large $t$.
\end{proposition}

{\it Remark:  Comparison with the results in \cite{PSE}.} Propositions \ref{comparison} and \ref{ppropp} shows the following behaviour of the decoherence function. In dimension $d=1$, for an infra-red behaviour of the coupling function $g(k)\sim |k|^{1/2}$, and for large times, the decoherence function is $\Gamma(t)\sim\xi(0)t$, which becomes $\Gamma(t)\sim Tt$, for small temperatures (see Remark 11 after \fer{i4}). This is the same result as obtained in \cite{PSE}. In dimension $d=3$, the contribution to the decoherence function which is quadratic in $\lambda$ vanishes, see Proposition \ref{comparison}. This is in accordance with the result of \cite{PSE} that decoherence is incomplete in three dimensions.

\medskip
We conclude this section with the analysis of the resonances for energy-conserving interactions. The next result (see Appendix \ref{prooflemmalso} for a proof) shows that the kernel of $L_0$ is invariant under energy-conserving interactions. In particular, the degeneracy of the eigenvalue zero is not lifted under perturbation. This means that there are no resonances bifurcating out of the origin.
\begin{proposition}[Zero resonances]
\label{funnylemma}
The kernels of $L_0$, $K_\lambda$ and $K_\lambda(\omega)$ coincide. They are spanned by the vectors
$$
\chi_0^{(s)} = \varphi_s\otimes\varphi_s\otimes\Omega_{\r,\beta},\ \ \ \ s=1,\ldots,N.
$$
The kernel of $(K_\lambda(\omega))^*$ is spanned by
$$
\widetilde\chi_0^{(s)} = \varphi_s\otimes\varphi_s\otimes(\bbbone + T_s)\Omega_{\r,\beta},\ \ \ \ s=1,\ldots,N,
$$
where $T_s=O(\lambda)$ satisfies $\scalprod{\Omega_{\r,\beta}}{T_s\Omega_{\r,\beta}}=O(\lambda^2)$. In particular, there are no resonances bifurcating from the origin.
\end{proposition}

As an application of this result, we show, using resonance theory, that the populations are constant (a fact we know already from Corollary \ref{corollary11}). Proposition \ref{funnylemma} gives us the following expresssion for the projection onto the zero resonances:
\begin{eqnarray}
Q_0&=& \frac{-1}{2\pi\i} \int_{\Gamma_0} (K_\lambda(\omega)-z)^{-1} \d z\label{51}\\
&=& \sum_{s=1}^N w_s^{-1} |\varphi_s\otimes\varphi_s\otimes\Omega_{\r,\beta}\rangle\langle \varphi_s\otimes\varphi_s\otimes (\bbbone+T_s)\Omega_{\r,\beta}|,
 \nonumber
\end{eqnarray}
where
$$
w_s= \scalprod{(\bbbone+T_s)\Omega_{\r,\beta}}{\Omega_{\r,\beta}} = 1+\av{T_s^*}_{\Omega_{\r,\beta}} =1+O(\lambda^2)
$$
is a normalization factor. Using \fer{51} in equation \fer{19} we obtain
\begin{eqnarray}
\av{\av{p_{n,m}}}_\infty &=&\delta_{n,m}\frac{\e^{-\beta E_m/2}}{\sqrt{Z_{\s,\beta}}} \scalprod{B^*\psi_0}{\varphi_m\otimes\varphi_m}\nonumber\\
&=&\delta_{n,m} \scalprod{(B^*\psi_0)\otimes\Omega_{\r,\beta}}{(p_{n,m}\otimes\bbbone_\s\otimes\bbbone_\r)\Omega_{\beta,0}}\nonumber\\
&=& \delta_{n,m}\av{p_{n,m}}_{t=0}.
\label{50}
\end{eqnarray}
To see how to recover the fact that the populations are constant in time we note that since $p_{n,n}\otimes\bbbone_\s\otimes\bbbone_\r$ commutes with $K_\lambda(\omega)$ (this is not true for $p_{n,m}$, $m\neq n$) and since $K_\lambda(\omega)\Omega_{\beta,0}=0$ we have
$$
(K_\lambda(\omega)-z)^{-1} (p_{n,n}\otimes\bbbone_\s\otimes\bbbone_\r)\Omega_{\beta,0} = -z^{-1} (p_{n,n}\otimes\bbbone_\s\otimes\bbbone_\r)\Omega_{\beta,0}.
$$
The contour integrals in \fer{17} can then be evaluated,
\begin{eqnarray*}
\av{p_{n,n}}_t -\av{p_{n,n}}_{t=0} &=& \av{p_{n,n}}_{t=0}\ \frac{1}{2\pi\i}\int_{\rx+\i\omega'/2}\!\frac{\e^{\i tz}}{z}\ \d z .
\end{eqnarray*}
The last integral has to be understood in the sense of the appropriate limit \fer{iri}. By a standard contour deformation we see that we may take $\omega'>0$ in this integral as large as we please, so the r.h.s. is zero (see the proof of Proposition \ref{propremainder} in Appendix \ref{prooflemmalso} for details). Consequently, $\av{p_{n,n}}_t =\av{p_{n,n}}_{t=0}$, i.e., $[\rhobar_t]_{n,n} = [\rhobar_0]_{n,n}$.

Next we examine the nonzero resonances. We have proven above that, under energy-conserving perturbations, there are no resonances bifurcating out of the origin. Now we show that nonzero eigenvalues may migrate into the upper complex plane or may stay on the real line, depening on the energy-conserving interaction. We know from \fer{50} that the ergodic means of the off-diagonal matrix elements tend to zero as $t\rightarrow\infty$. The off-diagonal matrix elements can be purely oscillatory, which corresponds to resonances staying on the real axis, or they decay in case the resonances move into the upper complex plane. To examine which of the two cases happens we consider the level shift operator $\Lambda_e$ associated to an eigenvalue $e$ of $L_\s$, given in \fer{lso1}. Since $G$ is diagonal, only the diagonal term \fer{20} is present. Let $I_e$ be the set of indices $m,n$ s.t. $e=E_m-E_n$, where the $E_j$ are the eigenvalues of $H_\s$. We have $P_e=\sum_{I_e}p_m\otimes p_n$, where $p_j=|\varphi_j\rangle\langle\varphi_j|$, and we obtain easily the relation
$$
\Lambda_{e} = \sum_{I_e} p_m\otimes p_n \left[ (\gamma_m^2-\gamma_n^2)\scalprod{g}{\omega^{-1}g}-\i\pi\xi(0)(\gamma_m-\gamma_n)^2\right],
$$
see also \fer{35} (where the integral is understood to be taken over $\rx^d$). This shows that if the coupling to the low energy modes of the reservoir is weak, $p>\frac{2-d}{2}$, then the imaginary part of all the level shift operators vanish, since $\xi(0)=0$. Note that for $e=0$ the sum extends over $p_m\otimes p_m$ and hence ${\mathrm Im}\Lambda_0 =0$, which is consistent with the result of Proposition \ref{prop7.1}. Furthermore, if $G$ is constant on ${\mathrm Ran}P_e$ then ${\mathrm Im}\Lambda_{e} = 0$ as well. In these cases the resonances stay on the real line, to second order in $\lambda$. Proposition \ref{prop7.1} asserts that this actually happens to all orders in $\lambda$, a result which one could extract from the resonance theory as well, by performing a deeper analysis. For instance, one sees from the explicit form of the interaction operator $I(\omega)$ (Appendix \ref{appA}, after \fer{aa6}) that 
\begin{equation}
I(\omega)\varphi_m\otimes\varphi_n\otimes\Omega_{{\rm R},\beta} = (\gamma_m-\gamma_n)\big(\bbbone_\s\otimes\bbbone_\s\otimes U(\omega) \varphi_\beta(g)\big)\varphi_m\otimes\varphi_n\otimes\Omega_{{\rm R},\beta},
\label{starstar}
\end{equation}
and consequently, that if $\gamma_m=\gamma_n$, then the Feshbach map applied to $\overline K_\lambda(\omega)$ is simply equal to $eP_e$. The isospectrality of the Feshbach map implies that the eigenvalue $e$ of $L_0$ does not move under the perturbation (to any order in $\lambda$).

\subsection{Observations about quantum registers}
\label{sectregisters}

Consider a register of $L$ qubits located at positions $x_j$, $j=1,\ldots,L$, which do not interact directly with each other. This $L$-bit register is placed in an environment, modelled by a thermal Bose field described in Section \ref{opensystsection}. It is known \cite{PSE,DG} that the decoherence properties of the register depend on the relation between the spacing of the qubits in the register and the correlation length of the reservoir. Consider the following two cases:
\begin{itemize}
\item[(i)] The qubits are very far apart, s.t. $\min_{i\neq j}|x_i-x_j|$ is much larger than the correlation length of the reservoir.
\item[(ii)] The qubits are packed very closely together, so that $\max_{i,j}|x_i-x_j|$ is much smaller than the correlation length of the reservoir.
\end{itemize}
In case (i) we expect that a good approximation to the true dymamics is given by a register where each qubit is coupled to its own reservoir. Such a system is described by the zero-temperature Hilbert space of pure states
$$
\hh_{\mathrm i}= \bigotimes_{j=1}^L \hh_{\s,j}\otimes\hh_{\r,j},
$$
where $\hh_{\s,j}$ and $\hh_{\r,j}$ are the Hilbert spaces of the $j$-th qubit and the $j$-th reservoir.  (Recall the notation of Sections \ref{sectintrodd} and \ref{opensystsection}.)  The subindex i indicates that the qubits are coupled to individual, independent reservoirs. The Hamiltonian is given by
\begin{equation}
H_{{\mathrm i},\lambda}= \sum_{j=1}^L H_{\lambda,j},
\label{hnc}
\end{equation}
where $H_{\lambda,j}$ acts nontrivially only on the $j$-th qubit-reservoir pair, and is given by \fer{ii1}. Accordingly, the positive temperature Hilbert space is $\h_{\mathrm i}=\otimes_{j=1}^L\h_{\s,j}\otimes\h_{\r,j}$, and the Liouville operator has the form $L_\lambda = \sum_{j=1}^L L_{\lambda,j}$, where each $L_{\lambda,j}$ generates the dynamics of a single qubit coupled to its own environment, as in Section \ref{intro1}. Let us label the state $\varphi_{m_1}\otimes\cdots\otimes\varphi_{m_L}$ of the register by ${\mathbf m}=(m_1,\ldots,m_L)\in \{0,1\}^L$. The reduced density matrix for the register, $\rhobar_t$, is simply the product of the reduced density matrices of the single qubits, $\rhobar_{j,t}$. Its matrix elements are
$$
[\rhobar_t]_{{\mathbf m},{{\mathbf n}}}=\prod_{j=1}^L [\rhobar_{j,t}]_{m_j,n_j},
$$
where each $[\rhobar_{j,t}]_{m_j,n_j}$ evolves according to \fer{i1}-\fer{i4}.

\medskip

Next let us consider the case (ii). We expect that the true dymamics is well approximated by an interaction term where all the qubits sit in the same location. In this situation one observes {\it collective decoherence}, \cite{PSE,DG}. Here, the subindex ``c'' stands for ``collective''. The Hilbert space of pure states (at zero temperature) is
$$
\hh_{\mathrm c} = \left(\bigotimes_{j=1}^L \hh_{\s,j}\right)\otimes\hh_\r,
$$
where $\hh_{\s,j}$ is the Hilbert spaces of the $j$-th qubit and $\hh_\r$ is that of the reservoir. The Hamiltonian is given by
\begin{equation}
H_{{\mathrm c},\lambda}= \sum_{j=1}^L H_{\s,j} +H_\r +\lambda \left(\sum_{j=1}^L G_j\right)\otimes\varphi(g),
\label{hc}
\end{equation}
where $H_{\s,j}$ is as above and where $G_j$ acts as a fixed matrix $G$ on the $j$-th qubit and trivially on all other qubits. The positive temperature Hilbert space is
$$
\h_{\mathrm c} = \left(\bigotimes_{j=1}^L \h_j\right)\otimes\h_\r,
$$
where $\h_j=\cx^2\otimes\cx^2$, and $\h_\r$ is given in \fer{90}. The generator of dynamics takes the form $L_{{\mathrm c},\lambda}=L_0+\lambda V_{\mathrm c}$, with
\begin{equation}
L_0 = \sum_{j=1}^L L_{\s,j} +L_\r \mbox{\ \ and \ \ } V_{\mathrm c}=\left(\sum_{j=1}^L G_j\otimes\bbbone_j\right)\otimes\varphi_\beta(g)
\label{r3}
\end{equation}
(compare also with \fer{r-1}-\fer{83} and \fer{73.1}). Here, $L_{\s,j}$ acts non-trivially, as $H_\s\otimes\bbbone - \bbbone\otimes H_\s$, only on the $j$-th qubit space, and  $G_j\otimes\bbbone_j$ acts as $G\otimes\bbbone$ on the $j$-th qubit space and trivially on the other qubits. Our general result, Theorems \ref{lemma} and \ref{prop3.2} and Proposition \ref{LSO} are valid for this model. However, the analysis of the level shift operators associated to the Liouville operator \fer{r3} becomes increasingly more involved with growing $L$. This is simply due to the size of the matrices representing the level shift operators. We point out, though, that the particular structure of $V_{\mathrm c}$ may facilitate the spectral analysis of the level shift operators. Furthermore, since we have explicit formulas for the levels shift operators, our method may be suitable for a computer-based analysis.

\medskip
The Hamiltonians $H_{{\mathrm i},\lambda}$ and $H_{{\mathrm c},\lambda}$, \fer{hnc} and \fer{hc}, are in a certain sense extreme cases, as described by (i) and (ii) above, of an intermediate Hamiltonian. The latter is given in \cite{PSE} as
\begin{equation}
H_\lambda = \sum_{j=1}^L H_{\s,j}\otimes \bbbone +\bbbone\otimes H_\r +\lambda \sum_{j=1}^L  G_j\otimes\varphi\left(e^{-\i kx_j}g\right).
\label{r1}
\end{equation}
Here, $g$ is a fixed form factor, typically imagined to be so that its inverse Fourier transform, $\check g$, is peaked around the origin in $x$ space. The Fourier transform of the function $\check g$ shifted to the position of the $j$-th qubit, $\check g_j(x):=\check g(x-x_j)$, is then just the $e^{-\i kx_j}g$ appearing in the interaction of \fer{r1}.

\medskip

The following are two problems in connection with quantum registers:
\begin{itemize}
\item[P1] Derive the form of Hamiltonian \fer{r1} from a real physical situation, e.g. where qubit $j$ is represented by two levels of an atom located at position $x_j$. Quantify the approximation schemes regarding  cases (i), (ii) above.

\item[P2] Analyze the level shift operators associated with the interaction \fer{r3}, analytically or numerically. Obtain the dynamics of reduced density matrix elements (as in \fer{i1}-\fer{i4}) for the collectively decohering quantum register. Identify coherent subspaces.
\end{itemize}

\begin{appendix}

\section{Unitary transformation of the positive-temperature Hilbert space}
\label{appgluing}

It is convenient to work with a unitarily transformed version of the Araki-Woods representation for the Bose field \fer{66}-\fer{67}, c.f. \cite{JP1,FM1}. The Hilbert space \fer{66} is transformed unitarily as
\begin{equation}
{\cal F}(L^2(\rx^3,\d^3k))\otimes {\cal F}(L^2(\rx^3,\d^3k)) \mapsto {\cal F}\equiv {\cal F}(L^2(\rx\times S^2,\d u\d\sigma)),
\label{90}
\end{equation}
according to the map
\begin{eqnarray}
\lefteqn{ a^*(f_1)\cdots a^*(f_m)\Omega\otimes  a^*(g_1)\cdots a^*(g_n)\Omega}\nonumber\\
& &\longmapsto a^*(\chi_+{\cal T}f_1)\cdots a^*(\chi_+{\cal T} f_m)a^*(\chi_-{\cal T} g_1)\cdots a^*(\chi_-{\cal T}g_n)\Omega,
\label{90.1}
\end{eqnarray}
where the vectors $\Omega$ are the vacua in the respective Fock spaces and, and where the $a^\#$ are the creation and annihilation operators in the respective Fock spaces. We have introduced ${\cal T}$ which maps functions $f(k)\in L^2(\rx^3,\d^3k)$ into functions $({\cal T}f)(u,\sigma)\in L^2(\rx\times S^2,\d u\d \sigma)$, according to
\begin{equation}
({\cal T}f)(u,\sigma)= u
\left\{
\begin{array}{ll}
f(u,\sigma) & \mbox{if $u\geq 0$},\\
-\e^{\i\phi} \overline f(-u,\sigma) & \mbox{if $u<0$},
\end{array}
\right.
\label{91}
\end{equation}
where $f$ is represented in polar coordinates and $\phi$ is an arbitrary real phase. This phase is a parameter which can be chosen appropriately to satisfy Condition (A) after \fer{7} for a given coupling function $g$, see \cite{FM1}. 
The $\chi_\pm$ in \fer{90.1} are indicator functions, $\chi_+(u)=1$ if $u\geq 0$, $\chi_+(u)=0$ if $u<0$, and $\chi_-=1-\chi_+$.

One verifies that the thermal annihilation operators, represented in the Araki-Woods representation by \fer{corr1}, take the following form in the unitarily transformed system:
\begin{equation}
a_\beta(f) = a\big(\sqrt{1+\mu_\beta(u)} \chi_+(u) u f(u,\sigma)\big) - a^*\big(\e^{\i\phi} \sqrt{\mu_\beta(-u)} \chi_-(u) u \overline f(-u,\sigma)\big).
\label{13}
\end{equation}
(The $a^*_\beta(f)$ are obtained by taking the adjoint on the r.h.s. of \fer{13}). A short calculation shows that the thermal field operator \fer{83} becomes, via the unitary transformation,
\begin{equation}
\varphi_\beta(f) = \frac{1}{\sqrt 2}(a_\beta^*(f)+a_\beta(f)) = \frac{1}{\sqrt 2}(a^*(f_\beta)+a(f_\beta))=:\varphi( f_\beta),
\label{14}
\end{equation}
for $f\in L^2(\rx^3)$, where $f_\beta$ is given in \fer{7}, and where the $\varphi$ in the r.h.s. is the field operator in $\cal F$.
The equilibrium state is represented by the vacuum vector of $\cal F$,
\begin{equation}
\Omega_{{\rm R},\beta}=\Omega.
\label{93}
\end{equation}
For a one-body operator $O$ acting on wave functions of the variables $(u,\sigma)$, we write 
\begin{equation}
\d\Gamma(O) = \int_{\rx\times S^2}  a^*(u,\sigma) O a(u,\sigma)\, \d u\d\sigma.
\label{xx}
\end{equation}
for the second quantization of the operator $O$. 
The new representation has the advantage that the dynamics of the field is generated simply by
\begin{equation}
L_\r = \d\Gamma(u),
\label{94}
\end{equation}
the second quantization of the operator of multiplication by $u$. We have $L_\r\Omega_{\r,\beta}=0$, and for $z\in\cx$,
\begin{equation}
\e^{z L_\r} \varphi_\beta(f)\e^{-zL_\r} = 2^{-1/2}\left( a_\beta\big( \e^{-\overline z u}f\big) + a_\beta^*\big(\e^{z u}f\big)\right),
\label{10}
\end{equation}
which gives the dynamics for $z=\i t$. 

It follows from \fer{73.1}, \fer{31} and \fer{14} that the Liouville operator $L_\lambda$ acting on $\h_\s\otimes{\cal F}$ is given by 
\begin{eqnarray}
L_\lambda &=& L_0 +\lambda V,\label{nr1}\\
L_0 &=& L_\s+L_\r = H_\s\otimes\bbbone_\s -\bbbone_\s\otimes H_\s + \d\Gamma(u),\label{nr2}\\
V &=& G\otimes\bbbone_\s\otimes \varphi(g_\beta). \label{nr3}
\end{eqnarray}

One verifies that the operator $D$, \fer{opD}, is represented in the unitarily transformed space as $\d\Gamma(\i\partial_u)$, 
it generates translations in the variable $u\in\rx$ (see also \cite{MMS1}). The unitary group $U(\omega)$, \fer{Uomega}, is thus given by the {\it translation group} 
\begin{equation}
U(\omega) = \e^{-\i\omega \d\Gamma(\i\partial_u)}.
\label{xxy}
\end{equation}
The spectrally deformed Liouville operator acting on $\h_\r\otimes{\cal F}$ is
\begin{equation}
L_\lambda(\omega) = L_0+\omega N +\lambda V(\omega),
\label{sdlop}
\end{equation}
where $N=\d\Gamma(\bbbone)$ is the number operator in $\cal F$, and where $V(\omega)=\e^{-\omega\d\Gamma(\partial_u)} V\e^{\omega\d\Gamma(\partial_u)}$ (see also \fer{aa5}).

Observables of $\r$ are operators that can be built from sums and products of thermal creation and annihilation operators. Strictly speaking, one considers bounded operators built from exponentiated thermal field operators, $\e^{\i \varphi_\beta(f)}$. They form the so-called {\it Weyl-algebra}, a von Neumann algebra 
\begin{equation}
\mm_\r\subset{\cal B}({\cal F}),
\label{vna}
\end{equation}
see e.g. \cite{BR}.

\section{The operators $K_\lambda$ and $K_\lambda(\omega)$}
\label{appA}

We consider the positive temperature Hilbert space in its form given in Appendix \ref{appgluing}. The operator $K_\lambda$ can be expressed in terms of the non-interacting Liouville operator $L_0$, the interaction $V$, see \fer{nr1}-\fer{nr3}, and {\it modular data} $J,\Delta$ (see e.g. \cite{BR}) associated to the vector $\Omega_{\beta,0}$ and the von Neumann algebra $\mm$, \fer{vna}. $J$ is an anti-unitary operator and $\Delta$ is a self-adjoint non-negative operator. The defining properties of $J$ and $\Delta$ are
\begin{equation}
J\Delta^{1/2} M\Omega_{\beta,0} = M^*\Omega_{\beta,0},
\label{defnJ}
\end{equation}
for any $M\in\mm$, where $M^*$ is the adjoint operator of $M$. From this property and the facts that $L_0\Omega_{\beta,0}=0$ and $V=V^*$ we readily see that the operator
\begin{eqnarray}
K_\lambda &=& L_0 + \lambda I, \label{4}\\
I&=& V - J\Delta^{1/2} V J\Delta^{1/2},
\label{4.1}
\end{eqnarray}
satisfies $K_\lambda\Omega_{\beta,0}=0$. The operators $J$ and $\Delta$ satisfy $J\Delta=\Delta^{-1}J$, and therefore $J\Delta^{1/2}VJ\Delta^{1/2} = J\Delta^{1/2}V\Delta^{-1/2}J$. The theory of von Neumann algebras (Tomita-Takesaki) tells us that conjugation with $\Delta^{1/2}$ leaves the algebra $\mm$ invariant (provided the operators in question exist; $\Delta$ is unbounded), and furthermore, that all operators of the form $J M J$, $M\in\mm$, commute with all operators $N\in\mm$, so the subtracted term in \fer{4} commutes with all observables of $\s+\r$ and hence {\it does not alter the dynamics}.

We now give explicit expressions for the operators $J$, $\Delta$ and $K_\lambda$, see also \cite{BR,FM1,MMS1,MMS2}. The modular data is
\begin{equation}
J=J_\s\otimes J_\r \mbox{\ \ and\ \ }
\Delta = \Delta_\s\otimes\Delta_\r,
\label{5.01}
\end{equation}
where
\begin{eqnarray}
\Delta_\s &=& \e^{-\beta L_\s},\label{5.1}\\
\Delta_\r &=& \e^{-\beta L_\r},\label{5.2}\\
J_\s\phi_l\otimes\phi_r &=& \cc\phi_r\otimes \cc\phi_l,\label{5.3}\\
J_\r\psi_n(u_1,\sigma_1,\ldots,u_n,\sigma_n) &=& \e^{\i n\phi}\overline\psi_n(-u_1,\sigma_1,\ldots,-u_n,\sigma_n),\label{5.4}
\end{eqnarray}
where the action of the antilinear operator $\cc$ is to take the complex conjugate of vector coordinates in the basis $\{\varphi_j\}_{j=1}^N$ of $\h_\s$, and $\overline\psi_n$ is the complex conjugate of $\psi_n\in{\cal F}$ (see \fer{90}). The phase $\phi\in\rx$ is the one appearing in \fer{7}. It may be chosen suitably to satisfy condition (A) (see after \fer{7}), given a form factor $g$. Relation \fer{5.4} shows that 
\begin{equation}
J_\r a^\#(f(u,\sigma)) J_\r = a^\#(\e^{\i\phi}\overline f(-u,\sigma)),
\label{12}
\end{equation}
for $f\in L^2(\rx\times S^2)$. 

We use relations \fer{31} and \fer{5.01}-\fer{5.3} to obtain
\begin{equation}
I = V - V',
\label{9}
\end{equation}
where 
\begin{eqnarray}
V' &=& \bbbone_\s\otimes\e^{-\frac\beta 2 H_\s}\overline G \e^{\frac\beta 2 H_\s}\otimes \frac{1}{\sqrt{2}} \left[ a^*\big(g_\beta\big)+ a\big(\e^{-\beta u}g_\beta \big) \right].\ \ \ \
\label{9.1}
\end{eqnarray}
We have set $\overline G=\cc G\cc$ here, and we recall that $g_\beta$ is defined in \fer{7}.

We now give the explicit form of the spectrally deformed operator $K_\lambda(\omega)=L_0+\omega N +I(\omega)$, where $U(\omega)=\e^{-\i\omega\d\Gamma(\i\partial_u)}$ (see also \fer{sdlop}). The transformation under $U(\omega)$ of creation and annihilation operators is given by
\begin{equation}
U(\omega) a^\#(f) U(\omega)^{-1} = a^\#(f(\cdot +\omega)),\ \ \ \omega\in\rx, 
\label{aa1}
\end{equation}
where $f(\cdot +\omega)$ is the shifted function $(u,\sigma) \mapsto f(u+\omega,\sigma)$. Relation \fer{aa1} can be written in the form $U(\omega)a^\#(f)U(\omega)^{-1} = a^\#(\e^{\omega\partial_u}f)$. In order to obtain an analytic extension of \fer{aa1} to complex $\omega$, we need to take the complex conjugate of $\omega$ in the argument of the annihilation operator (since the latter is anti-linear in its argument). We thus have (see also \fer{14})
\begin{eqnarray}
I(\omega) &=& V(\omega) -V'(\omega),\label{aa4} \\
V(\omega) &=&  G\otimes\bbbone_\s\otimes\frac{1}{\sqrt{2}}\left[ a^*(g_\beta(\cdot+\omega)) + a(g_\beta(\cdot+\overline\omega))\right],\label{aa5} \\
V'(\omega) &=& \bbbone_\s\otimes\e^{-\frac\beta 2 H_\s}\overline G \e^{\frac\beta 2 H_\s}\otimes \frac{1}{\sqrt{2}} \left[ a^*(g_\beta(\cdot +\omega)) + a\big(\e^{-\beta (u+\overline\omega)}g_\beta(\cdot+\overline\omega) \big)\right].\ \ \ \ \ \ \ \ \label{aa6}
\end{eqnarray}
Let us finally prove the validity of \fer{starstar}. Definition \fer{xxy} of $U(\omega)$ implies that 
$$
U(\omega)\varphi_m\otimes\varphi_n\otimes\Omega_{\r,\beta} = \varphi_m\otimes\varphi_n\otimes\Omega_{\r,\beta}.
$$
Further, it follows from \fer{5.1}-\fer{5.4} that 
$$
J\Delta^{1/2}\varphi_m\otimes\varphi_n\otimes\Omega_{\r,\beta}=\e^{-\beta(E_m-E_n)/2} \varphi_n\otimes\varphi_m\otimes\Omega_{\r,\beta}.
$$
Therefore, we have 
\begin{eqnarray}
\lefteqn{
V'(\omega)\varphi_m\otimes\varphi_n\otimes\Omega_{\r,\beta}}\nonumber \\
 &=&  U(\omega) V'\varphi_m\otimes\varphi_n\otimes\Omega_{\r,\beta}\nonumber \\
&=&\e^{-\beta(E_m-E_n)/2} U(\omega)J\Delta^{1/2} V\varphi_n\otimes\varphi_m\otimes\Omega_{\r,\beta}\nonumber \\
&=&\gamma_n \e^{-\beta(E_m-E_n)/2}  U(\omega)J\Delta^{1/2} (\bbbone_\s\otimes\bbbone_\s\otimes\varphi_\beta(g))\varphi_n\otimes\varphi_m\otimes\Omega_{\r,\beta}\nonumber \\
&=&\gamma_n U(\omega)(\bbbone_\s\otimes\bbbone_\s\otimes\varphi_\beta(g))\varphi_m\otimes\varphi_n\otimes\Omega_{\r,\beta},
\label{aa2}
\end{eqnarray}
where we have also used in the third step that $G\varphi_n=\gamma_n\varphi_n$, and in the last step that $J_\r\Delta_\r\varphi_\beta(g)\Omega_{\r,\beta} = \varphi_\beta(g)\Omega_{\r,\beta}$ (which follows from \fer{defnJ} and the fact that $\varphi_\beta(g)$ is self-adjoint). Combining \fer{aa2} with
$$
V(\omega)\varphi_m\otimes\varphi_n\otimes\Omega_{\r,\beta} = \gamma_m U(\omega)(\bbbone_\s\otimes\bbbone_\s\otimes\varphi_\beta(g))\varphi_m\otimes\varphi_n\Omega_{\r,\beta}
$$
and with the fact that $I(\omega) = V(\omega)-V'(\omega)$, we obtain \fer{starstar}.

\section{Perturbation of equilibrium states and outline of proof of Theorem \ref{cor1}}
\label{AppCor}
We give an outline of the expansion of the equilibrium state $\rho(\beta,\lambda)$  of a coupled system (or its reduction to a subsystem) in terms of the coupling constant $\lambda$. Explicit calculations for the situation of the qubit  yield a proof of Theorem \ref{cor1}. We do not carry them out here.

Consider an $N$-level system coupled to the Bose field, as described in Sections \ref{opensystsection}, \ref{intro1}. It is well known \cite{BR,BFS,FM1,JP1,Mlso,MMS1,MMS2} that the equilibrium state  $\rho(\beta,\lambda)$ w.r.t. the interacting dynamics, \fer{72}, is represented by the vector
\begin{equation}
\Omega_{\beta,\lambda}=\frac{\e^{-\beta L_\lambda/2}\Omega_{\beta,0}}{\|\e^{-\beta L_\lambda/2}\Omega_{\beta,0}\|} \in \cal H.
\label{omegabeta}
\end{equation}
Here, the operator $L_\lambda$ and the equilibrium state of the uncoupled system, $\Omega_{\beta,0}$, are given by \fer{73.1} and \fer{3.1}, respectively, and $\cal H$ is the Hilbert space \fer{73}. Our task is to expand the average 
\begin{equation}
[\rhobar_\infty]_{m,n}=\scalprod{\Omega_{\beta,\lambda}}{(p_{n,m}\otimes\bbbone_\s)\Omega_{\beta,\lambda}}
\label{mmm1}
\end{equation}
in powers of $\lambda$, where we recall the definition $p_{n,m}=|\varphi_n\rangle\langle\varphi_m|$, the $\{\varphi_j\}$ forming an orthonormal basis diagonalizing the Hamiltonian $H_\s$. Since $L_0\Omega_{\beta,0}=0$, it is convenient to use the Dyson series,
\begin{eqnarray*}
\e^{-\beta L_\lambda/2}\Omega_{\beta,0} &=& \e^{-\beta L_\lambda/2}\e^{\beta L_0/2}\Omega_{\beta,0}\\
&=& \Omega_{\beta,0} +\sum_{n=1}^\infty (-\lambda)^n\int_0^{\beta/2}\d s_1\cdots\int_0^{s_{n-1}}\d s_n V(s_1)\cdots V(s_n)\Omega_{\beta,0},
\end{eqnarray*}
where $V(s):= \e^{-s L_0}V\e^{s L_0}$. Accordingly, it is clear how to arrive at an expansion of $\Omega_{\beta,\lambda}$, \fer{omegabeta}, and hence of the averages \fer{mmm1}, in powers of $\lambda$. \hfill $\blacksquare$

\section{Proofs of propositions}
\label{prooflemmalso}

{\bf Proof of Proposition \ref{propremainder}.\ }
We expand the resolvent in \fer{17.1} in $\lambda$,
\begin{equation}
(K_\lambda(\omega)-z)^{-1} = (K_0(\omega)-z)^{-1} - \lambda (K_0(\omega)-z)^{-1}I(\omega) (K_\lambda(\omega)-z)^{-1}.
\label{reseq}
\end{equation}
The contribution to $R$ coming from the free resolvent $(K_0(\omega)-z)^{-1}$ is given by
\begin{equation}
\frac{-1}{2\pi\i}\int_{\rx+\frac{\i}{2}[\mu+\omega'/2]}  f(z) \d z,
\label{17.2}
\end{equation}
where $f(z):=\e^{\i tz} \scalprod{B^*\psi_0}{(L_\s-z)^{-1} A \Omega_{\s,\beta}}$, the inner product being that of $\h_\s$.
To arrive at \fer{17.2} we use that $A$ is an observable of $\s$, that $\Omega_{\beta,0}=\Omega_{\s,\beta}\otimes\Omega_{\r,\beta}$ and that $(L_\r+\omega N)\Omega_{\r,\beta}=0$. The integral is understood in the sense of \fer{iri}. Our first goal is to show that \fer{17.2} is actually zero. Consider the integral $\int_{-a}^b f(x+\frac{\i}{2}[\mu+\omega'/2]) \d x$.
Since the only singularities of the integrand are poles on the real axis we can deform the contour of integration, yielding that for any $r>\frac 12[\mu+\omega'/2]$,
\begin{eqnarray}
\lefteqn{ \int_{-a}^b f(x+\textstyle\frac{\i}{2}[\mu+\omega'/2]) \d x=}\label{17.3}\\
&& \int_{-a}^b f(x+\i r) \d x +\int_{\frac 12[\mu+\omega'/2]}^r f(-a+\i y) \d y -\int_{\frac 12[\mu+\omega'/2]}^r f(b+\i y) \d y.\nonumber
\end{eqnarray}
It is easy to see that $|f(-a+\i y)|< C\e^{-ty}/a$ for $a$ sufficiently large, and $|f(b+\i y)|< C\e^{-ty}/b$ for $b$ sufficiently large. It follows from \fer{17.3} that
\begin{equation}
\int_{-a}^b f(x+\textstyle \frac{\i}{2}[\mu+\omega'/2]) \d x= \int_{-a}^b f(x+\i r) \d x +O(1/a)+O(1/b),
\label{17.4}
\end{equation}
for $a,b\rightarrow\infty$, and where the remainder terms are uniform in $r$. In the limit $r\rightarrow\infty$ the first integral on the r.h.s. of \fer{17.4} vanishes, so by taking first $r\rightarrow\infty$ and then $a,b\rightarrow\infty$ we see that \fer{17.4} and thus \fer{17.2} are both zero.

Consequently, the term of order $\lambda^0$ in \fer{reseq} does not contribute to the integral in \fer{17.1}. By iterating the resolvent equation we see that all terms with odd powers in $\lambda$ do not contribute either, because the interaction $I$ is linear in creation and annihilation operators, and we take a ``vacuum'' expectation in \fer{17.1}. Thus we have
\begin{eqnarray}
\lefteqn{ R=\frac{-\lambda^2}{2\pi\i}\int_{\rx+\frac{\i}{2}[\mu+\omega'/2]}  \e^{\i tz} \big\langle(B^*\psi_0)\otimes\Omega_{\r,\beta}, (K_0(\omega)-z)^{-1}I(\omega)\times}\label{17.10}\\
&&\ \ \ \ \ \ \ \ \ \  \ \ \ \ \ \ \ \ \ \ \ \ \ \ \ \ \  \times (K_\lambda(\omega)-z)^{-1}I(\omega) (K_0(\omega)-z)^{-1} A \Omega_{\beta,0}\big\rangle \d z.\nonumber
\end{eqnarray}
Finally we want to show that the last integral is $O(\e^{- \frac t2[\mu+\omega'/2]})$. It is not hard to see that the norm of the integrand is bounded above by
$$
C \e^{-\frac t2[\mu+\omega'/2]} \|[(K_0(\omega)-z)^*]^{-1}\Phi_1\|\, \| (K_0(\omega)-z)^{-1}\Phi_2\|,
$$
for some constant $C$ and where $\Phi_1=(B^*\psi_0)\otimes\Omega_{\r,\beta}$ and $\Phi_2=A \Omega_{\beta,0}$. In order to conclude that $R=O(\lambda^2\e^{-\frac t2[\mu+\omega'/2]})$ it thus suffices to show that
\begin{equation}
\int_{\rx+\frac{\i}{2}[\mu+\omega'/2]} \|(K_0(\omega)-z)^{-1}\Phi_j\|^2 \d z<\infty.
\label{17.11}
\end{equation}
The integrand in \fer{17.11} is $\scalprod{\Phi_j}{[(L_0-x)^2 + (\omega' N-\frac 12[\mu+\omega'/2])^2]^{-1}\Phi_j}$,
which is readily seen to be integrable w.r.t. $x\in\rx$ (using for instance the spectral theorem for the commuting self-adjoint operators $L_0$ and $N$). We have thus shown that $R=O(\lambda^2\e^{-\frac t2[\mu+\omega'/2]})$.

We point out that we did not use the specific form of $\Phi_1$ in these estimates. The present argument works for all vectors $\Phi_1$ corresponding to any initial state $\rho_0$ on $\h_\s\otimes\h_\r$. \hfill $\blacksquare$

\medskip

{\bf Proof of Proposition \ref{LSO}.\ }
This is an easy calculation using the following explicit form of all operators involved, as presented in Appendix A.

\medskip

{\bf Proof of Proposition \ref{comparison}.\ } We refer to the paragraph after Proposition \ref{funnylemma} for a proof of the fact that the populations are independent of time. We now concentrate on the off-diagonals. By using that $G$ and $H_\s$ commute, it is not hard to see that the Feshbach map applied to $K_\lambda(\omega)$,
$$
F_{P_e,z}(K_\lambda(\omega)) = P_e \big( e -\lambda^2 I(\omega)\overline P_{\!e} (\overline{K}_\lambda(\omega)-z)^{-1}\overline P_{\!e} I(\omega)\big) P_e,
$$
is diagonal in the basis $\{\varphi_m\otimes\varphi_n\otimes \Omega_{\r,\beta}\}$ of ${\mathrm Ran} P_e$, where $m$ and $n$ are indices s.t. $e=E_m-E_n$. By the reconstruction formula for eigenvectors, \fer{28}, it follows that the eigenvectors of $K_\lambda(\omega)$ are of the form $\varphi_m\otimes\varphi_n\otimes (\bbbone+ T_{m,n})\Omega_{\r,\beta}$, where the $T_{m,n}$ are operators on $\h_\r$ satisfying $T_{m,n}=O(\lambda)$ and $\scalprod{\Omega_{\r,\beta}}{T_{m,n}\Omega_{\r,\beta}}=O(\lambda^2)$. (Compare also with Proposition \ref{funnylemma} and its proof.) In a similar way we see that the eigenvectors of $[K_\lambda(\omega)]^*$ are of the form $\varphi_m\otimes\varphi_n\otimes (\bbbone+\widetilde T_{m,n})\Omega_{\r,\beta}$, for some operators $\widetilde T_{m,n}$ on $\h_\r$ satisfying $\widetilde T_{m,n}=O(\lambda)$ and $\scalprod{\Omega_{\r,\beta}}{\widetilde T_{m,n}\Omega_{\r,\beta}}=O(\lambda^2)$.
Let $d(\varepsilon_e^{(s)})$ be the degeneracy of the resonance energy $\varepsilon_e^{(s)}$. Then we have
\begin{eqnarray}
\lefteqn{Q_e^{(s)}=
\sum_{j=1}^{d(\varepsilon_e^{(s)})} w_{m_{s,j},n_{s,j}}^{-1}}\label{d-d}\\
&& \times  |\varphi_{m_{s,j}}\otimes \varphi_{n_{s,j}}\otimes (\bbbone+T_{m_{s,j},n_{s,j}})\Omega_{\r,\beta}\rangle\langle \varphi_{m_{s,j}}\otimes \varphi_{n_{s,j}}\otimes (\bbbone+\widetilde T_{m_{s,j},n_{s,j}})\Omega_{\r,\beta}|,
\nonumber
\end{eqnarray}
where the normalization weights are
$$
w_{m_{s,j},n_{s,j}} := \scalprod{(\bbbone+\widetilde T_{m_{s,j},n_{s,j}})\Omega_{\r,\beta}}{(\bbbone+T_{m_{s,j},n_{s,j}})\Omega_{\r,\beta}} = 1+O(\lambda^2).
$$
We can now use expression \fer{d-d} in expansion \fer{24} to arrive at
\begin{equation}
[\rhobar_t]_{m,n} = \e^{\i t\varepsilon_{E_n-E_m}^{(s(m,n))}} C_{m,n}(\psi_0) + O(\lambda^2\e^{-\omega' t/2}),
\label{q1}
\end{equation}
where
\begin{eqnarray}
\varepsilon_{E_n-E_m}^{(s(m,n))} &=& E_n-E_m-{\textstyle \frac 12}\lambda^2  (\gamma_n^2-\gamma_m^2)\scalprod{g}{\omega^{-1}g}\label{q2}\\
&& +{\textstyle \frac 12}\i\lambda^2(\gamma_n-\gamma_m)^2\lim_{\epsilon\downarrow 0}\scalprod{g}{\coth\left(\frac{\beta\omega}{2}\right)\frac{\epsilon}{\omega^2+\epsilon^2}g} +O(\lambda^4).
\nonumber
\end{eqnarray}
The $C_{m,n}(\psi_0)$ in \fer{q1} are given by the following expression (see also the calculation leading to \fer{50})
\begin{eqnarray}
\lefteqn{ C_{m,n}(\psi_0) }\nonumber\\
&=& \scalprod{(B^*\psi_0)\otimes\Omega_{\r,\beta}}{\varphi_n\otimes\varphi_m\otimes(\bbbone+T_{n,m})\Omega_{\r,\beta}} \frac{\e^{-\beta E_m/2}}{\sqrt{Z_{\s,\beta}}}\  \frac{1+\langle\widetilde T_{n,m}^*\rangle_{\Omega_{\r,\beta}}}{w_{n,m}}\nonumber\\
&=& [\rhobar_0]_{m,n} (1+O(\lambda^2)). \label{q3}
\end{eqnarray}
Relations \fer{q1}, \fer{q2} and \fer{q3} show assertion \fer{q4} of the proposition, where
\begin{eqnarray*}
\lefteqn{
\delta_{E_n-E_m}:= -{\textstyle \frac 12}\lambda^2  (\gamma_n^2-\gamma_m^2)\scalprod{g}{\omega^{-1}g}}\\
&& +{\textstyle \frac 12}\i\lambda^2(\gamma_n-\gamma_m)^2\lim_{\epsilon\downarrow 0}\scalprod{g}{\coth\left(\frac{\beta\omega}{2}\right)\frac{\epsilon}{\omega^2+\epsilon^2}g}.
\end{eqnarray*}
 \hfill $\blacksquare$

\medskip

{\bf Proof of Proposition \ref{prop7.1}.\ }
We absorb the coupling constant $\lambda$ into the matrix $G$ (rescale $G$ so that $\lambda=1$). The reduced density matrix is given by
$$
\overline \rho_t = \tr_\r\left[ \e^{-\i tL } ( \rhobar_0\otimes|\Omega_{\r,\beta}\rangle\langle\Omega_{\r,\beta}| )  \e^{\i tL}\right],
$$
where the trace is taken over $\h_\r$, \fer{66} with $L^2(\rx^3,\d^3k)$ replaced by $L^2(\rx^d,\d^dk)$, and where $L=H_\s+ L_\r+ G\otimes\varphi_\beta(g)$ acts on the Hilbert space $\h=\cx^N\otimes \h_\r$. Thus we have
\begin{equation}
[\rhobar_t]_{m,n}=[\rhobar_0]_{m,n} \e^{-\i t(E_m-E_n)} \ \omega_{\r,\beta}\left( \e^{-\i t(\gamma_n\varphi_\beta(g)+L_\r)}\e^{\i t(\gamma_m\varphi_\beta(g)+L_\r)}\right),
\label{b1}
\end{equation}
where we denote the equilibrium state of $\r$ by $\omega_{\r,\beta}(\cdot) =\tr_\r(|\Omega_{\r,\beta}\rangle\langle\Omega_{\r,\beta}|\, \cdot\, )$. Now we apply the Trotter product formula,
$$
\omega_{\r,\beta}\left( \e^{-\i t(\gamma_n\varphi_\beta(g)+L_\r)}\e^{\i t(\gamma_m\varphi_\beta(g)+L_\r)}\right) =
\lim_{M\rightarrow\infty} \omega_{\r,\beta}\left( [X_n(M)]^M[X_m(M)^*]^M\right),
$$
where $X_n(M):=\e^{-\i tL_\r/M}\e^{-\i t\gamma_n\varphi_\beta(g)/M}$. Using that
\begin{eqnarray*}
\lefteqn{
[X_n(M)]^M [X_m(M)^*]^M}\\
&& =[X_n(M)]^{M-1}W_\beta\left(-\frac{\i t}{M}(\gamma_n-\gamma_m) \e^{-\i \omega t/M}g\right) [X_m(M)^*]^{M-1},
\end{eqnarray*}
where $W_\beta(f):=\e^{\i\varphi_\beta(f)}$ is the thermal Weyl operator, and using the relation
$$
W_\beta(f)W_\beta(g) = \e^{-\frac \i2 {\mathrm Im}\scalprod{f}{g}}W_\beta(f+g),
$$
we obtain by induction the formula
\begin{equation}
[X_n(M)]^M [X_m(M)^*]^M =\exp\left\{-\frac\i 2\sum_{K=1}^M S_K\right\} W(g_M),
\label{b2}
\end{equation}
where
$
g_M = -\frac tM(\gamma_n-\gamma_m)\sum_{k=1}^K \e^{-\i k\omega\frac t M}
$
and
$$
S_K = \frac{t^2}{M^2}(\gamma_n-\gamma_m)(\gamma_n+\gamma_m)\sum_{k=1}^{K-1} {\mathrm Im}\scalprod{g}{\e^{-\i k\omega\frac t M}g}.
$$
It follows from \fer{b2} and $\omega_{\r,\beta}(W_\beta(f))=\e^{-\frac14\scalprod{f}{\coth(\beta\omega/2)f}}$ (see e.g. \cite{BR}) that in the limit $M\rightarrow\infty$ (where the sums over $M$ turn into easy integrals which can be evaluated explicitly), we get
\begin{eqnarray}
\lefteqn{
\lim_{M\rightarrow\infty} \omega_{\r,\beta}\left( [X_n(M)]^M[X_m(M)^*]^M\right) }\nonumber\\
&&= \exp\left[-\i(\gamma_n-\gamma_m)(\gamma_n+\gamma_m)S(t) -(\gamma_n-\gamma_m)^2\Gamma(t)\right],
\label{b3}
\end{eqnarray}
with $S(t)$ and $\Gamma(t)$ defined in Proposition \ref{prop7.1}. The proof of this proposition is now completed by combining \fer{b1} and \fer{b3}. \hfill $\blacksquare$

\medskip
{\bf Proof of Proposition \ref{funnylemma}.\ }
We first notice that the Feshbach map \fer{29} for $e=0$ and with spectral parameter $z=0$ vanishes,
\begin{equation}
F_{P_0,0}(K_\lambda(\omega))= -\lambda^2 P_0 I(\omega)\overline P_0\big(\overline K_\lambda(\omega)\big)^{-1}\overline P_0 I(\omega)  P_0 =0.
\label{fl2}
\end{equation}
This is a simple consequence of the facts that ${\mathrm Ran} P_0$ is spanned by $\{\varphi_s\otimes\varphi_s\otimes\Omega_{\r,\beta}\}$ and that
\begin{equation}
I(\omega)\ \varphi_s\otimes\varphi_s\otimes\Omega_{\r,\beta} = 0.
\label{fl1}
\end{equation}
To see \fer{fl1}, simply use \fer{4}, \fer{4.1}, \fer{9} and \fer{9.1}, and that
$
J_\s\Delta_\s^{1/2} (G\otimes\bbbone_\s) J_\s\Delta_\s^{1/2}=\bbbone_\s\otimes G,
$
which holds since $[H_\s,G]=0=[\Delta_\s,G]$ and $G$ is self-adjoint.

Relation \fer{fl2} implies that $\dim{\mathrm Ker}K_\lambda =N$. To obtain a basis for ${\mathrm Ker}K_\lambda$ we use the reconstruction formula \fer{28} and notice that, again due to \fer{fl1}, only the term with $P_0$ survives.

Proceeding as in the proof of Lemma \ref{littlelemma} it is readily seen that
$$
F_{P_0,0}([K_\lambda(\omega)]^*)=[F_{P_0,0}(K_\lambda(\omega))]^*=0,
$$
so $\dim{\mathrm Ker}[K_\lambda(\omega)]^*=N$ and we reconstruct a basis for ${\mathrm Ker}[K_\lambda(\omega)]^*$ using \fer{28},
$$
\widetilde\chi_0^{(s)} = \left[ P_0 -\lambda\overline P_0 ([\overline K_\lambda(\omega)]^*)^{-1}\overline P_0 [I(\omega)]^* P_0 \right]\varphi_s\otimes\varphi_s\otimes\Omega_{\r,\beta}.
$$
Since all the operators commute with the spectral projections of $L_\s$ we may ``pull'' $\varphi_s\otimes\varphi_s$ through the operator $[\cdots]$ to the left, and it is easy to identify the $T_s$ having the properties given in the proposition. Note though that equation \fer{fl1} is not correct if we replace $I(\omega)$ by $[I(\omega)]^*$, so $T_s\neq 0$.
\hfill $\blacksquare$

\end{appendix}

\end{document}